\documentclass[3p,12pt]{elsarticle}

\usepackage{lineno,hyperref}
\usepackage{graphicx,subfloat,enumerate}
\usepackage{amssymb}
\usepackage{bm,amsmath}
\usepackage[caption=false]{subfig}
\usepackage{caption,color}
\usepackage{makecell}
\usepackage{subeqnarray}
\usepackage{multirow}
\usepackage{lscape}
\usepackage{rotating}
\usepackage{graphics}
\usepackage{epstopdf}
\usepackage{soul}
\usepackage{xcolor}
\setstcolor{red}

\journal{Elsevier}

\begin{document}
\title{Boosting the convergence of low-variance DSMC by GSIS}
\author{Liyan Luo}
\author{Qi Li \corref{cor1}}
\ead{liq33@sustech.edu.cn}
\author{Lei Wu \corref{cor1}}
\ead{wul@sustech.edu.cn}
\cortext[cor1]{Corresponding author}
\address{Department of Mechanics and Aerospace Engineering,
	Southern University of Science and Technology, Shenzhen 518055, China}
\date{\today}

\begin{abstract}
	
The low-variance direct simulation Monte Carlo (LVDSMC) is a powerful method to simulate low-speed rarefied gas flows. However, in the near-continuum flow regime, due to limitations on the time step and spatial cell size, it takes plenty of time to find the steady-state solution. Here we remove these deficiencies by coupling the LVDSMC with the general synthetic iterative scheme (GSIS) which permits the simulation at the hydrodynamic scale rather than the much smaller kinetic scale. As a proof of concept, we propose the stochastic-deterministic coupling method  based on the Bhatnagar-Gross-Krook kinetic model. First, macroscopic synthetic equations are derived exactly from the kinetic equation, which not only contain the Navier-Stokes-Fourier constitutive relation, but also encompass the higher-order terms describing the rarefaction effects. Then, the high-order terms are extracted from LVDSMC and fed into synthetic equations to predict macroscopic properties which are closer to the steady-state solution than LVDSMC. Finally, the state of simulation particles in LVDSMC is updated to reflect the change of macroscopic properties. As a result, the convergence to steady state is greatly accelerated, and the restriction on cell size and the time step are removed:
after simulating several canonical rarefied gas flows, we demonstrate that the LVDSMC-GSIS reduces the computational cost by two orders of magnitude in the near-continuum flow regime.	
\end{abstract}

\maketitle

\section{Introduction}

In many modern engineering problems, e.g., the space re-entry capsule, microelectromechanical systems and shale gas extraction, multiscale gas flows that span a wide range of Knudsen number are frequently encountered, and thus accurate/efficient simulation methods are urgently needed. At the macroscopic level, the Navier-Stokes-Fourier (NSF) equations, which are the pillars in traditional computational fluid dynamics, provide a mathematical model incorporating the linear constitutive relations, such as the Newton's law for stress and the Fourier's law for heat conduction. However, they are only valid in the flows when the characteristic length scale is much larger than the mean free path of gas molecules, i.e., when the Knudsen number is small. When the Knudsen number is large, the Boltzmann equation has to be adopted, which provides a unified kinetic approach from continuum flow to free molecular flow~\cite{cercignani-2000}. 

One of the most widely used methods to model the rarefied gas flow is the direct simulation Monte Carlo (DSMC) method. Although DSMC does not solve the Boltzmann equation directly~\cite{bird-1970}, it has been proven that its solution converges to that of the Boltzmann equation, when the number of simulation particles tends to infinity~\cite{wagner-1992,li-2013}. DSMC prevails due to the following advantages: it is unconditionally stable; it naturally captures the discontinuity of velocity distribution function; it is convenient to add complicated physical and chemical processes without increasing the computational cost significantly~\cite{bird-1994}.
However, there are some difficult issues  when using DSMC method in real applications. First, since DSMC is a stochastic method, the macroscopic information sampled from the simulation particles is inevitably subject to fluctuations. Several remedies are proposed to reduce  the fluctuation, for instances, the moment-guided Monte Carlo method~\cite{degond-2010} that matches the kinetic solution to deterministic solutions of moment equations, and the variance reduction technique~\cite{baker-2005B,homolle-2007B,Radtke-2011} that simulates only the deviational part from the equilibrium. Second, due to the splitting of advection and collision, the spatial cell size and time step should be smaller than the mean free path and mean collision time of gas molecules, respectively~\cite{bird-1994}, which lead to slow convergence and high computational cost for near-continuum flows. To reduce the computational cost, the asymptotic preserving DSMC method~\cite{pareschi-2000,pareschi-2001} and the NS-DSMC hybrid method~\cite{patronis-2013,stephani-2013} are developed, which partly solve these problems. 

On the contrary, the deterministic methods to solve the Boltzmann equation have made remarkable achievements over the past few decades. Based on the discrete velocity method (DVM), the unified gas-kinetic scheme (UGKS) has been developed~\cite{xu-2010B,huang-2012,Huang-2013}. The convection and collision are simultaneously solved, thus the restrictions on cell size and time step are removed. When the discrete scale is smaller than the kinetic scale, it has the same mechanism as traditional DVM. For discrete scale much larger than kinetic scale, the scheme is asymptotically approaching to Navier-Stokes equation, making it efficient to deal with multiscale problems~\cite{zhu-2016}. However,  for near-continuum flows, UGKS still needs many iterations to obtain the steady-state solution. To overcome this challenge, the general synthetic iterative scheme (GSIS) alternately solve the steady-state macroscopic synthetic equation and mesoscopic gas kinetic equation~\cite{SIS,CWF}, which can  not only find the steady-state solutions within dozens of iterations at any Knudsen number, but also uses a larger cell size than UGKS~\cite{FCA}.

Although the existing mature deterministic methods can be used to simulate many multiscale problems accurately and efficiently, they are still inferior to the DSMC method in dealing with hypersonic flows with chemical or physical process, where the Boltzmann collision operators are very complicated. Therefore, it is desirable to have a scheme mingling the advantages of both deterministic and stochastic methods.
In recent years, the stochastic particle methods based on Bhatnagar-Gross-Krook~\cite{fei-2020, fei-2021} model and Fokker-Planck model~\cite{Pfeiffer-2016, Jenny-2010, Gorji-2011, Gorji-2014} have been proposed. By simplifying the collision process, the computational efficiency in the continuum regimes has been improved compared to the original DSMC.
The former preserves the Navier-Stokes limit based on the BGK model~\cite{Bhatnagar-1954,Tumuklu-2016}, which is demonstrated to have high-order accuracy in space and time in the continuum regime, while in the latter method, a time integration scheme is applied, which is demonstrated to be more efficient than DSMC. 
The unified gas kinetic wave-particle method uses the wave-particle description to recover the non-equilibrium gas distribution function~\cite{Liu-2019, Zhu-2019, Li-2020}, where the particles without collisions are selectively sampled, while the unsampled particles are calculated by the deterministic method in the next evolution process. In the continuum regime, only a few non-collision particles are sampled and the evolution is dominated by the deterministic method. Thus, this method is efficient for high-speed multiscale problems. 

Since in rarefied gas flows the steady-state solution is frequently needed, it is not necessary to follow the time evolution of the kinetic equation. If there exists a scheme that directly pull/guide the solution to the final steady state, then the computational efficiency will be further improved on top of the above-mentioned methods. The GSIS servers this purpose. Inspired by the success of GSIS in multiscale problems, based on the Boltzmann equation and simplified kinetic models, the similar technique is expected to accelerate the slow convergence of DSMC method in the simulation of near-continuum flows. As a proof of concept, we couple the GSIS with the low-variance (LV) DSMC  to achieve asymptotic preservation and fast convergence in all flow regimes. If this is successful, then the speed-up of nonlinear DSMC by GSIS will be straightforward.

A brief outline is sketched below. The linearized Boltzmann equation with BGK operator is introduced in Section~\ref{sec_description}; the synthetic equations are derived and the GSIS-LVDSMC coupling method is proposed in Section~\ref{sec_gsis}; the accuracy and efficiency of the proposed method are assessed in canonical rarefied gas flows in Sections~\ref{sec_1d} and~\ref{sec_fourier}. Finally,  conclusions as well as future perspectives are summarized in Section~\ref{sec_conclusion}.

\section{The linearized BGK equation}\label{sec_description}

The BGK equation is widely used because of its simplicity:
\begin{equation}
\frac{\partial f}{\partial t} + \bm{c} \cdot \frac{\partial f}{\partial \bm{x}} + \frac{\partial (\bm{a}f)}{\partial \bm{c}}=  \frac{p}{\mu} (f_{loc}-f),
\label{eq:BTE}
\end{equation}
where $f(t, \bm{x}, \bm{c})$ is the velocity distribution function, with $t$ being the time, $\bm{x}$ the spatial coordinates, and $\bm{c}$ the molecular velocity; $\bm{a}$ is the external acceleration, $p$ is the gas pressure, $\mu$ is the shear viscosity of the gas, and 
\begin{equation}\label{eq:M-B}
f_{loc} = \frac{n}{{\left(2\pi R T\right)}^{3/2}}\exp\left( -\frac{| \bm{c}-\bm{u} |^{2}}{2R T}\right)
\end{equation}
is the local equilibrium distribution function, with $n$, $\bm{u}$, $T$, and $R$  being the number density, flow velocity, temperature, and gas constant of the gas, respectively. 

Introducing the dimensionless variables $\hat{f}=c_0^{3}f/n_0$,  $\hat{t}=c_0 t/L$, $\bm{\hat{x}}=\bm{x}/L$, $\bm{\hat{c}}=\bm{c}/c_0$, and $\bm{\hat{a}}=\bm{a}L/c_0^2$, where $n_0$ is the reference number density, 
$c_0=\sqrt{2RT_0}$ is the most probable speed at the reference temperature $T_0$, $L$ is the characteristic flow length, the BGK equation is normalized to the following form:
\begin{equation}\label{eq:DBE}
\frac{\partial \hat{f}}{\partial \hat{t}} + \bm{\hat{c}} \cdot \frac{\partial \hat{f}}{\partial \bm{\hat{x}}} + \frac{\partial (\bm{\hat{a}} \hat{f})}{\partial \bm{\hat{c}}}=  \delta_{rp}\left(\hat{f}_{loc}-\hat{f}\right),
\end{equation}
where $\delta_{rp}$ is the rarefaction parameter defined as (it is related to the Knudsen number $\text{Kn}$ as $\delta_{rp}=\frac{\sqrt{\pi}}{2\text{Kn}}$): 
\begin{equation}
\delta_{rp}=\frac{pL}{\mu c_0}.
\end{equation}

In the linearized kinetic model, the velocity distribution function can be written as the combination of the equilibrium distribution function $\hat{f}_0=\pi^{-3/2}\exp\left(-\bm{\hat{c}}^2\right)$ at the reference state and the perturbed distribution function $\hat{h}$:
\begin{equation}
\hat{f}(\hat{t},\bm{\hat{x}},\bm{\hat{c}})=\hat{f}_0(\bm{\hat{c}})+\alpha\hat{h}(\hat{t},\bm{\hat{x}},\bm{\hat{c}}), 
\label{eq:LBE_f}
\end{equation} 
where the small constant $\alpha$ is related to the amplitude of perturbation, with $\alpha\hat{h}$ satisfying $|\alpha\hat{h}/\hat{f}_0|\ll 1$. However, the distribution function $\hat{h}$ is not necessarily smaller than the fixed equilibrium distribution function $\hat{f}_0$. With this in mind, Eq.~\eqref{eq:DBE} can be linearized as~\cite{SIS}:
\begin{equation}
\begin{aligned}[b]
\frac{\partial \hat{h}}{\partial \hat{t}}+\hat{\bm{c}}\cdot\frac{\partial \hat{h}}{\partial \hat{\bm{x}}}=L_{BGK}\left(\hat{h},\hat{f}_{0}\right)
\underbrace{-\frac{\partial (\bm{\hat{a}}\hat{f}_0)}{\partial \bm{\hat{c}}}}_{S},\\
L_{BGK}=\delta_{rp}\left[\rho+2\bm{\hat{c}}\cdot\bm{\hat{u}}+\left(\hat{c}^2-\frac{3}{2}\right)\tau\right]\hat{f}_0-\delta_{rp}\hat{h},
\end{aligned}
\label{eq:LBEP}
\end{equation}
where $\rho$, $\bm{\hat{u}}$ and $\tau$ are the perturbed dimensionless number density, flow velocity and temperature, respectively. Note that the normalized acceleration is also small so that the system permits linearization; the corresponding term is usually treated as the source term $S$. For instances, in the Poiseuille flow and thermal transpiration, due to the small pressure gradient and temperature gradient in the $x_3$ direction, the ``equivalent'' source terms are
\begin{equation}
S=\begin{cases}
\hat{c}_3\hat{f}_0, \quad &\text{Poiseuille flow},\\
\left(\frac{5}{2}-\hat{c}^2\right)\hat{c}_3\hat{f}_0, \quad &\text{Thermal transpiration.}
\end{cases}
\label{eq:source}
\end{equation}

The dimensionless macroscopic quantities (which are further normalized by the constant $\alpha$) are defined as the moments of the perturbed velocity distribution function: 
\begin{equation}\label{eq:gai}
\begin{aligned}[b]
\rho=\int \hat{h}d^3\hat{c}, \quad \hat{\bm{u}}=\int \hat{\bm{c}}\hat{h}d^3\hat{c}, \quad \tau=\frac{2}{3}\int \left(\hat{c}^2-\frac{3}{2}\right)\hat{h}d^3\hat{c}, \\ 
\hat{\sigma}_{ij}=2\int\left(\hat{c}_i\hat{c}_j-\frac{\hat{c}^2}{3}\delta_{ij}\right)\hat{h}d^3\hat{c}, \quad
\hat{\bm{q}}=\int \left(\hat{c}^2-\frac{5}{2}\right)\hat{\bm{c}}\hat{h}d^3\hat{c},  
\end{aligned}
\end{equation}
where $\hat{\sigma}_{ij}$ and $\hat{\bm{q}}$ are the dimensionless deviatoric stress and heat flux, respectively; $\delta_{ij}$ is the Kronecker delta function, and the subscripts $i,j=1,2,3$ indicate directions in the Cartesian coordinate system.

\section{The coupling of LVDSMC and GSIS}\label{sec_gsis}


Although the velocity distribution function $f$ is non-negative, the perturbed distribution function $\hat{h}$ can be positive or negative. Therefore, in LVDSMC, by introducing the positive and negative deviational particles, the perturbed distribution function over a single computational cell can be expressed as~\cite{Radtke-2009}:
\begin{equation}
\hat{h}=N_{\text{eff}}\sum_{p\in \text{cell}}s_{p}\delta\left(\hat{\bm{x}}_{p}-\hat{\bm{x}}\right)\delta\left(\hat{\bm{c}}_p-\hat{\bm{c}}\right),
\label{eq:fd}
\end{equation}
where $N_{\text{eff}}$ is the number of gas molecules represented by one simulated particle, $s_p=\pm 1$ indicates the signs of positive and negative particles, and the subscript $p$ indicates the information of single simulation particle. The above equation shows that the $p$-th simulation particle in the cell has a position $\hat{\bm{x}}_{p}$ and velocity $\hat{\bm{c}}_{p}$. Combined with Eq.~\eqref{eq:gai}, the local macroscopic properties can be sampled over each computational cell (with volume $V_{\text{cell}}$) as: 
\begin{equation}\label{eq:after}
\begin{aligned}[b]
&\rho=\frac{N_{\text{eff}}}{V_{\text{cell}}}\sum_{p\in \text{cell}} s_p, 
\quad
\hat{\bm{u}}=\frac{N_{\text{eff}}}{V_{\text{cell}}}\sum_{p\in \text{cell}} s_p\hat{\bm{c}}_p,
\quad
\tau = \frac{2}{3}\frac{N_{\text{eff}}}{V_{\text{cell}}}\sum_{p\in \text{cell}} s_p\left(\hat{c}^2_p-\frac{3}{2}\right),
\\
&\hat{\sigma}_{ij}=2\frac{N_{\text{eff}}}{V_{\text{cell}}}\sum_{p\in \text{cell}} s_p\left[(\hat{c}_i)_p(\hat{c}_j)_p-\frac{\hat{c}^2_p}{3}\delta_{ij}\right],
\quad
\hat{\bm{q}}=\frac{N_{\text{eff}}}{V_{\text{cell}}}\sum_{p\in \text{cell}} s_p\hat{\bm{c}}_p\left(\hat{c}^2_p-\frac{5}{2}\right).
\end{aligned}
\end{equation}

The LVDSMC is a stochastic method, therefore, macroscopic properties outputted from LVDSMC are the cumulative average of each time step after the flow field reaches the steady-state, see the flowchart in Fig.~\ref{flowchart} without the blue box. It is efficient when the Knudsen number is large. However, due to the restriction on cell size and time step, the computational cost is large when the Knudsen number is small, albeit the variance reduction is achieved by simulating only the deviation from equilibrium. First, the total evolution (iteration) steps required to find the steady-state solution at least scale as $1/\text{Kn}^2$. Second, the number of collision pairs in each cell and time step increases with the number density, and thus leads to not only more computational time on collision subroutine, but also more deviational particles.

In order to increase the efficiency of LVDSMC, the evolution steps should be reduced significantly. Since we are interested in the steady-state solution, a scheme that guide the LVDSMC evolution directly to the steady state, without considering the intermediate evolution, is highly desired. Furthermore, to remove the restriction on cell size, the scheme should also asymptotically preserve the Navier-Stokes limit, so that the hydrodynamic scale (i.e., characteristic flow length), which is much large than the kinetic scale (i.e., mean free path), can be used. The recently developed GSIS perfect meets both requirements~\cite{FCA,CWF}, but has only been successfully applied to DVM-GSIS, which is a deterministic-deterministic coupling; this kind of coupling requires many discrete velocities hence many computational memory and time in hypersonic flow simulations. In order to reduce the computational memory and time, here we explore the possibility of stochastic-deterministic coupling scheme called GSIS-LVDSMC. Were this is successful, then the coupling of GSIS-DSMC is straightforward.


According to Ref.~\cite{CWF}, the essential idea of GSIS is that macroscopic synthetic equations are exactly derived from the kinetic equation, which not only contains the Navier-Stokes constitutive relations, but also encapsulates high-order terms to capture the rarefaction effects. On respectively multiplying Eq.~\eqref{eq:LBEP} with 1, $2\hat{c}_i$, and $\hat{c}^2-\frac{3}{2}$, and integrating with respect to the molecular velocity space, we have 
\begin{equation}
\begin{aligned}[b]
\frac{\partial \hat{u}_i}{\partial \hat{x}_i}=\int Sd^3\hat{c}, \\
\frac{\partial \rho}{\partial \hat{x}_i}+\frac{\partial \tau}{\partial \hat{x}_i}+\frac{\partial \hat{\sigma}_{ij}}{\partial \hat{x}_j}=\int 2\hat{c}_iSd^3\hat{c}, \\
\frac{\partial \hat{q}_i}{\partial \hat{x}_i}+\frac{\partial \hat{u}_i}{\partial \hat{x}_i}=\int \left(\hat{c}^2-\frac{3}{2}\right)Sd^3\hat{c}.
\end{aligned}
\label{eq:Evolution}
\end{equation}

Since the stress and heat flux are not closed, we again consider their governing equations by multiplying Eq.~\eqref{eq:LBEP} with $2 \left(\hat{c}_i\hat{c}_j-\frac{\delta_{ij}}{3}\hat{c}^2\right)$ and $\hat{c}_i\left(\hat{c}^2-\frac{5}{2}\right)$, respectively, and integrate with respect to the molecular velocity space, resulting 
\begin{equation}\label{eq:stress}
\underbrace{2\int\left(\hat{c}_i\hat{c}_j-\frac{\delta_{ij}}{3}\hat{c}^2\right)\hat{\bm{c}}\cdot\frac{\partial\hat{h}}{\partial\hat{\bm{x}}}d^3\hat{c}-2\frac{\partial \hat{u}_{<i}}{\partial \hat{x}_{j>}}}_{\text{HoT}_{\sigma_{ij}}}+\underbrace{2\frac{\partial \hat{u}_{<i}}{\partial \hat{x}_{j>}}=-\delta_{rp}\hat{\sigma}_{ij}}_{\rm{Newton's\ law}}
+\int 2 \left(\hat{c}_i\hat{c}_j-\frac{\delta_{ij}}{3}\hat{c}^2\right) Sd^3\hat{c},
\end{equation}
and 
\begin{equation}
\underbrace{\int\hat{c}_i\left(\hat{c}^2-\frac{5}{2}\right)\hat{\bm{c}}\cdot\frac{\partial\hat{h}}{\partial\hat{\bm{x}}}d^3\hat{c}-\frac{5}{4}\frac{\partial\tau}{\partial\hat{x}_i}}_{\text{HoT}_{q_{i}}}+\underbrace{\frac{5}{4}\frac{\partial\tau}{\partial\hat{x}_i}=-\delta_{rp}\hat{q}_i}_{\rm{Fourier's\ law}}
+ \int \hat{c}_i\left(\hat{c}^2-\frac{5}{2}\right)Sd^3\hat{c}.
\label{eq:heatflux}
\end{equation}
Note that in Eqs.~\eqref{eq:stress} and~\eqref{eq:heatflux}, the high-order terms $\text{HoT}_{\sigma_{ij}}$ and $\text{HoT}_{q_{i}}$ are computed directly from the perturbed velocity distribution functions, thus no approximations are introduced. Inevitably, the high-order terms evaluated from the stochastic method at each time step is subject to significant fluctuations, which may lead to numerical instability. In order to reduce the fluctuations to some extent, the time-averaged values of the high-order terms are adopted.
Also note that the velocity gradients in  Eq.~\eqref{eq:stress}  and temperature gradients in Eq.~\eqref{eq:heatflux} can not be canceled, since the one in the high-order term is statistically sampled from LVDSMC before solving the synthetic equations, while the one in Newton/Fourier's law will be solved from the synthetic equations, in order to guide the evolution of velocity and temperature in LVDSMC. The boundary conditions associated with the synthetic equations are extracted from LVDSMC. After obtaining the macroscopic flow properties from the synthetic equations, the deviational particles will be added or deleted accordingly, so that Eq.~\eqref{eq:after} can be satisfied by the updated local properties. Since the steady state synthetic equations with the Newton's law of stress and Fourier's law of heat conduction lead to diffusion-type of equations for the flow velocity and temperature, the information of updated deviational particles will reach the steady state much faster than the evolution of pure mesoscopic equations, so that the fast convergence can be achieved by coupling LVDSMC and GSIS. The flowchart of GSIS-LVDSMC algorithm is visualized in Fig.~\ref{flowchart}, and summarized as follows:

\begin{figure}[t]
	\centering
	\includegraphics[width=0.9\textwidth]{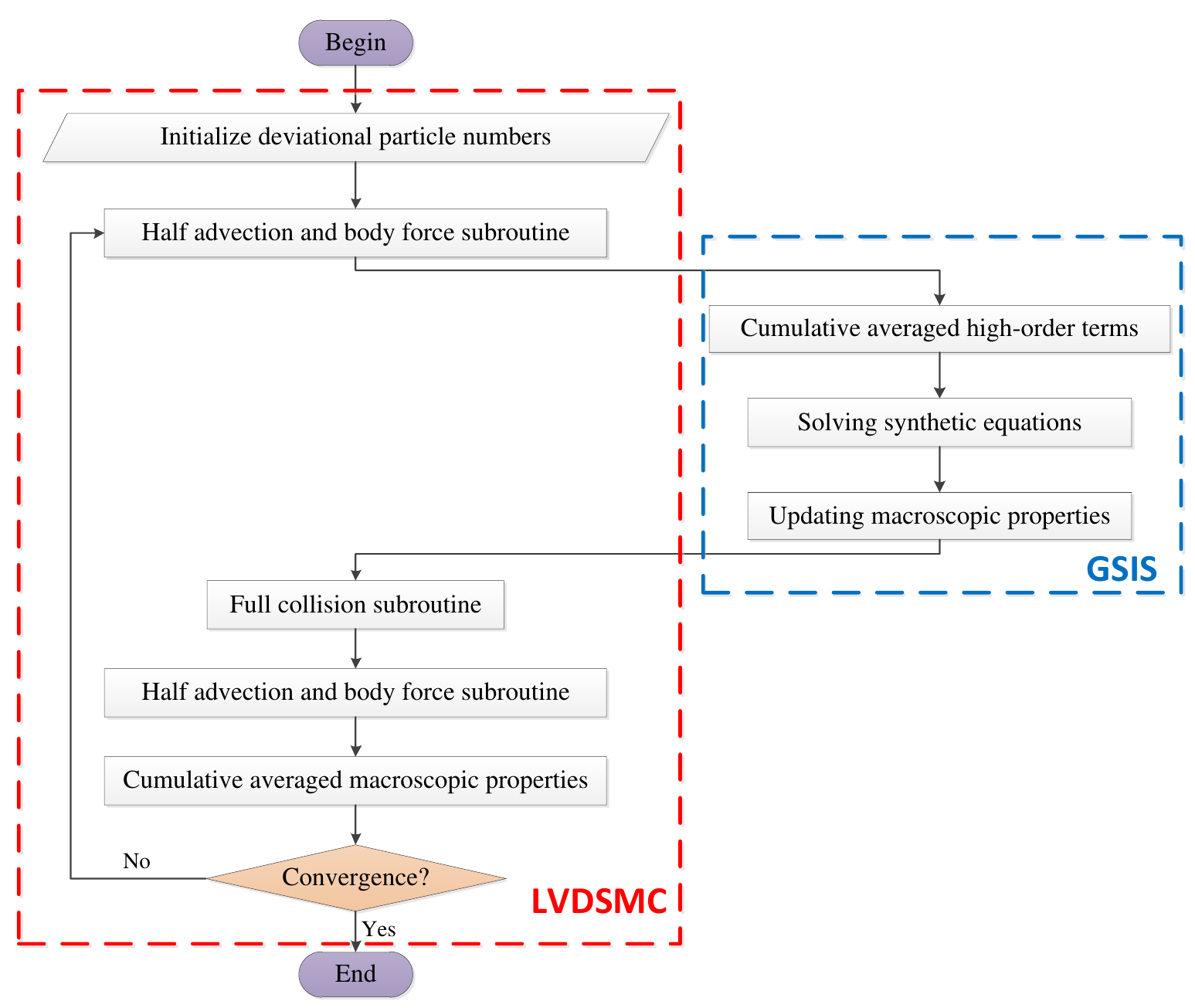}
	\caption{Flowchart of the GSIS-LVDSMC coupling algorithm. The red dashed box represents the original LVDSMC, and the blue dashed box indicates the additional part of GSIS.}
	\label{flowchart}
\end{figure}

\begin{enumerate}

	\item System initialization with local Maxwellian distributions (deviational particles number $= 0$);
	
	\item Half advection and linearized body force step as they are in the original LVDSMC~\cite{baker-2005B};
	
	\item Coupling of GSIS for fast convergence: 
	\begin{itemize}
		\setlength{\itemsep}{0pt}
		\setlength{\parsep}{0pt}
		\setlength{\parskip}{0pt}
		\item[\romannumeral1.] Calculate the time-averaged high-order terms $\text{HoT}_{\sigma_{ij}}$ and $\text{HoT}_{q_{i}}$ based on  Eqs.~\eqref{eq:stress} and~\eqref{eq:heatflux} in all cells, sample the macroscopic properties that need to be used as boundary conditions in synthetic equations;
		\item[\romannumeral2.] Solve the synthetic macroscopic equations \eqref{eq:Evolution} with constitutive relations~\eqref{eq:stress} and~\eqref{eq:heatflux}, and boundary conditions sampled from deviational particles;
		\item[\romannumeral3.] Update the macroscopic properties $\rho$, $\hat{\bm{u}}$ and $\tau$ by adding/deleting deviational particles so that Eqs.~\eqref{eq:after} can be satisfied.
	\end{itemize}
	\item Full collision step as in the original LVDSMC;
	
	\item Another half advection and linearized body force step (repeat step 2);
	
	\item Sample and accumulate the macroscopic properties based on Eq.~\eqref{eq:after}, and calculate the relative error of time-averaged values;
	
	\item Repeat steps 2-6 until the convergence criteria are met;
\end{enumerate}

\section{Numerical tests for the Poiseuille flow and thermal transpiration}\label{sec_1d}

We assess the accuracy and efficiency of GSIS-LVDSMC in the Poiseuille flow and thermal transpiration, from the continuum to free-molecular flow regimes. The mechanism of convergence-boosting is discussed in terms of the convergence rate, size cell, and time step.

\subsection{Algorithm validation and efficiency assessment}

It is noted from the source terms~\eqref{eq:source} of the Poiseuille flows and thermal transpiration that the velocity distribution function $\hat{h}$ in the steady state is an odd function of $\hat{c}_3$. Thus, the perturbed density and temperature become zero. Therefore, only the flow velocity has to be solved from the synthetic equation to boost convergence. To this end, the moment equation in Eq.~\eqref{eq:Evolution} is reduced to
\begin{equation}
\frac{\partial\hat{\sigma}_{13}}{\partial\hat{x}_1}+\frac{\partial\hat{\sigma}_{23}}{\partial\hat{x}_2}=\int 2\hat{c}_3Sd^3\hat{c}
=\begin{cases}
1, \quad &\text{Poiseuille flow},\\
0, \quad &\text{Thermal transpiration},
\end{cases}
\label{eq:stress_13and23}
\end{equation}
and the constitutive relation \eqref{eq:stress} of the shear stress remains unchanged in the presence of the source term~\eqref{eq:source}:
\begin{equation}
\hat{\sigma}_{i3}=-\delta_{rp}^{-1}\left[\int2\hat{c}_i\hat{c}_3\left(\hat{c}_1\frac{\partial\hat{h}}{\partial\hat{x}_1}+\hat{c}_2\frac{\partial\hat{h}}{\partial\hat{x}_2}\right)d^3\hat{c}-\frac{\partial\hat{u}_3}{\partial\hat{x}_i}+\frac{\partial\hat{u}_3}{\partial\hat{x}_i}\right], \quad (i=1,2).
\label{eq:stress_13and23_deduction}
\end{equation}

Substituting Eq.~\eqref{eq:stress_13and23_deduction} into Eq.~\eqref{eq:stress_13and23}, the diffusion-type synthetic equation for the flow velocity can be expressed as:
\begin{equation}
\begin{aligned}[b]
\frac{\partial^2\hat{u}_3}{\partial\hat{x}_1^2}+\frac{\partial^2\hat{u}_3}{\partial\hat{x}_2^2}&=-\delta_{rp}\int 2\hat{c}_3Sd^3\hat{c}-\frac{1}{4}\left(\frac{\partial^2F_{2,0,1}}{\partial\hat{x}_1^2}+\frac{\partial^2F_{1,1,1}}{\partial\hat{x}_1\partial\hat{x}_2}+\frac{\partial^2F_{0,2,1}}{\partial\hat{x}_2^2}\right)\\
&=\left\{
	\begin{aligned}
		-\delta_{rp}&-\frac{1}{4}\left(\frac{\partial^2F_{2,0,1}}{\partial\hat{x}_1^2}+2\frac{\partial^2F_{1,1,1}}{\partial\hat{x}_1\partial\hat{x}_2}+\frac{\partial^2F_{0,2,1}}{\partial\hat{x}_2^2}\right), \qquad \text{Poiseuille flow}, \\  
		&-\frac{1}{4}\left(\frac{\partial^2F_{2,0,1}}{\partial\hat{x}_1^2}+2\frac{\partial^2F_{1,1,1}}{\partial\hat{x}_1\partial\hat{x}_2}+\frac{\partial^2F_{0,2,1}}{\partial\hat{x}_2^2}\right), \qquad \text{Thermal transpiration}.
	\end{aligned}
\right.
\end{aligned}
\label{eq:SIS}
\end{equation}
where the high-order terms are defined and statistically sampled from LVDSMC as follows:
\begin{equation}
\begin{aligned}[b]
\frac{1}{4}F_{2,0,1} &=\int \hat{c}_3\left(2\hat{c}_1^2-1\right)\hat{h} d^3\hat{c}= \frac{N_{\text{eff}}}{V_{\text{cell}}}\sum_{p\in{\text{cell}}}s_p\hat{c}_3\left(2\hat{c}_1^2-1\right), \\
\frac{1}{4}F_{0,2,1} &=\int \hat{c}_3\left(2\hat{c}_2^2-1\right)\hat{h} d^3\hat{c}= \frac{N_{\text{eff}}}{V_{\text{cell}}}\sum_{p\in{\text{cell}}}s_p\hat{c}_3\left(2\hat{c}_2^2-1\right),\\
\frac{2}{4}F_{1,1,1} &=\int 4\hat{c}_1\hat{c}_2\hat{c}_3\hat{h} d^3\hat{c}= \frac{N_{\text{eff}}}{V_{\text{cell}}}\sum_{p\in{\text{cell}}}4s_p\hat{c}_1\hat{c}_2\hat{c}_3.
\label{eq:HOT}
\end{aligned}
\end{equation}
Meanwhile, the flow velocity $\hat{u}_3$ in the cells adjacent to walls are sampled from the simulation particles, which are used as the boundary conditions in solving the macroscopic equation~\eqref{eq:SIS}.

\subsubsection{1D cases}

In 1D simulation of the Poiseuille flow and thermal transpiration, the half spatial region $0\le\hat{x}_1\le0.5$ is uniformly divided into $N$ cells. The symmetric condition is applied at $\hat{x}_1=0.5$; the wall at $\hat{x}_1=0$ is fully diffuse, i.e., in this problem the velocity distribution function for particles entering into the simulation domain is zero.  

The synthetic equation~\eqref{eq:SIS} can be transformed into the following simple form:
\begin{equation}
\frac{\partial^2}{\partial\hat{x}_1^2}\left(\hat{u}_3+\frac{1}{4}F_{2,0,1}\right)\equiv\frac{\partial^2\Phi}{\partial\hat{x}_1^2}=\begin{cases}-\delta_{rp}, \quad &\text{Poiseuille flow,} \\ 0, \quad &\text{Thermal transpiration,}\end{cases}
\label{eq:fai}
\end{equation}
which is numerically solved by the central difference scheme with the same spatial discretization as in the LVDSMC:
\begin{equation}
	\frac{\partial^2\Phi_j}{\partial\hat{x}_1^2}=\frac{\Phi_{j-1}-2\Phi_{j}+\Phi_{j+1}}{\left(\Delta \hat{x}_1\right)^2}, \quad j=2,...,N.
\label{eq:CD}
\end{equation}
In the first cell, $\Phi_1$ is the combination of the local velocity sampled at each time step and the time-averaged high-order terms, which is sampled from LVDSMC and used as the boundary condition in solving Eq.~\eqref{eq:CD}. Due to symmetry, values in the $N$-th cell are the same as those in the $N+1$ virtual cell: $\Phi_{N}=\Phi_{N+1}$.

\begin{figure}[t]
	\centering
	\includegraphics[width=0.45\textwidth]{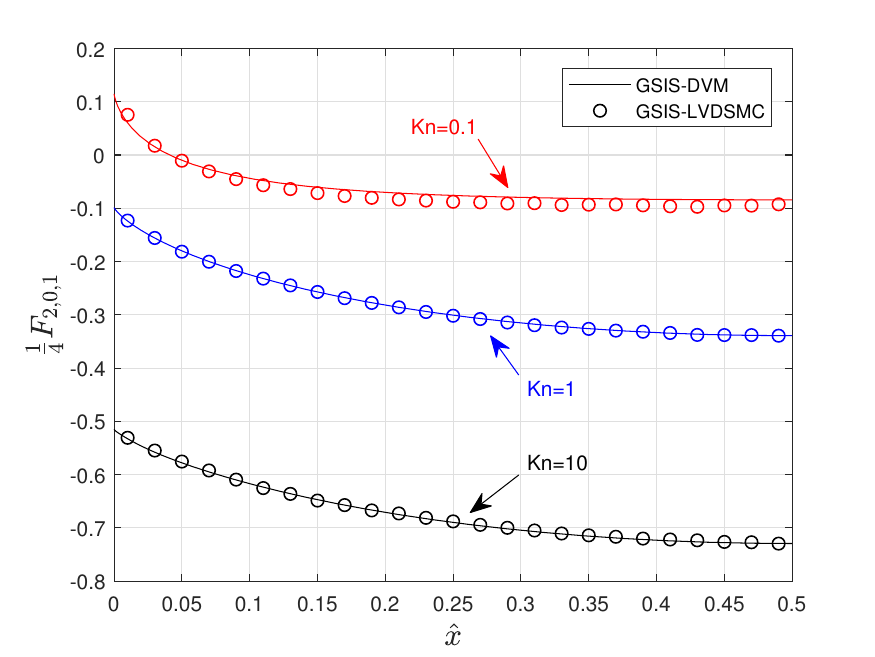}
	\includegraphics[width=0.45\textwidth]{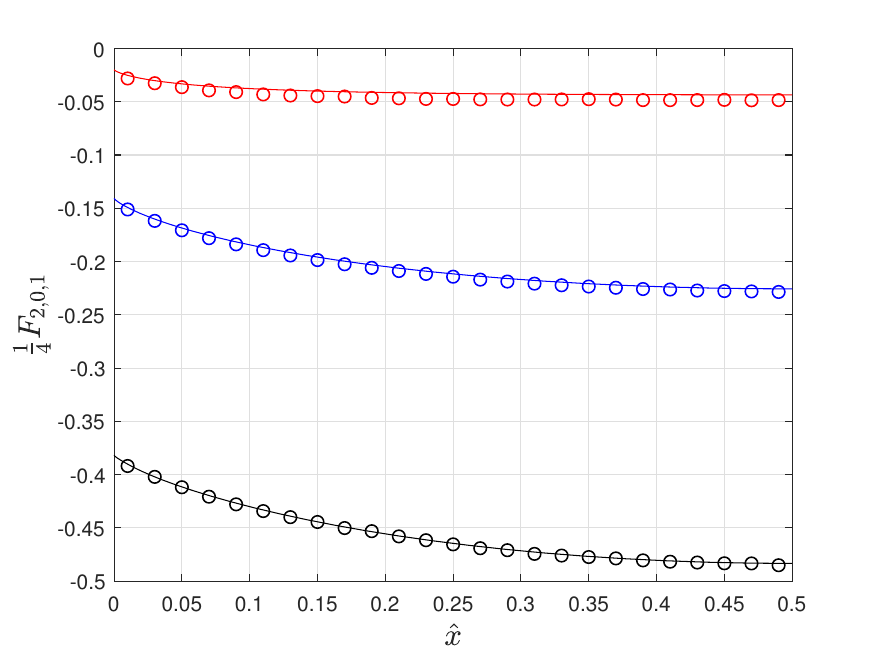}
	\caption{High-order term $\frac{1}{4}F_{2,0,1}$ obtained from GSIS-LVDSMC and GSIS-DVM~\cite{SIS} at different Knudsen number. (Left) Poiseuille flow. (Right) thermal transpiration. Note that the solution from the GSIS-DVM is chosen as the reference solution.  }
	\label{HO}
\end{figure}

The relative error in flow velocity between two successive steps, $k$ and $k+1$, is recorded during the simulation:
\begin{equation}
	\varepsilon = \int \left| \frac{\hat{u}_3^{(k+1)}}{\hat{u}_3^{(k)}}-1 \right| d\hat{x}_1.
	\label{eq:relative_error_1D}
\end{equation}
In the transition state, the velocity $\hat{u}_3$ is the time-averaged value sampled from the beginning of  simulation, and the system is regarded as reaching steady state when $\varepsilon<10^{-3}$. After that, the time averaging of all macroscopic values is restarted, and the simulation is terminated when $\varepsilon<10^{-6}$. We set a maximum number of steps $N_{\text{step,max}}$ to stop the simulations, even if the convergence criterion is not met, meaning that the simulation cannot be finished within an acceptable computational time.

Since the high-order terms~\eqref{eq:HOT} are essential to accurately capture the rarefaction effects, we compare their values obtained from GSIS-LVDSMC to the reference values from GSIS-DVM~\cite{SIS}. Figure~\ref{HO} shows good agreements in $F_{2,0,1}$:  only tiny deviations occur when $\text{Kn}=0.1$, due to the relative significant fluctuations of the particle method in the GSIS-LVDSMC algorithm.

\begin{figure}[t]
	\centering
	\includegraphics[width=0.45\textwidth]{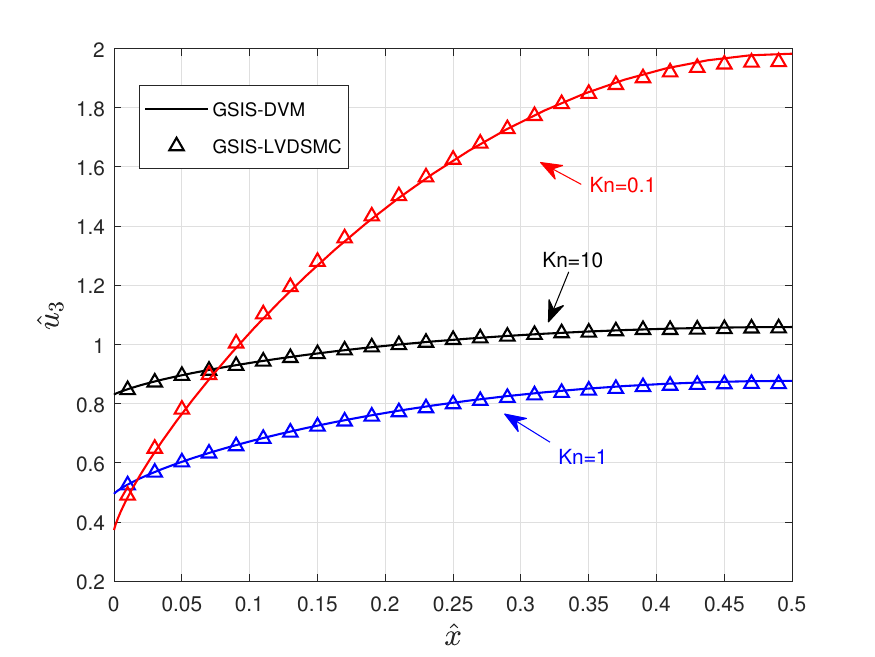}
	\includegraphics[width=0.45\textwidth]{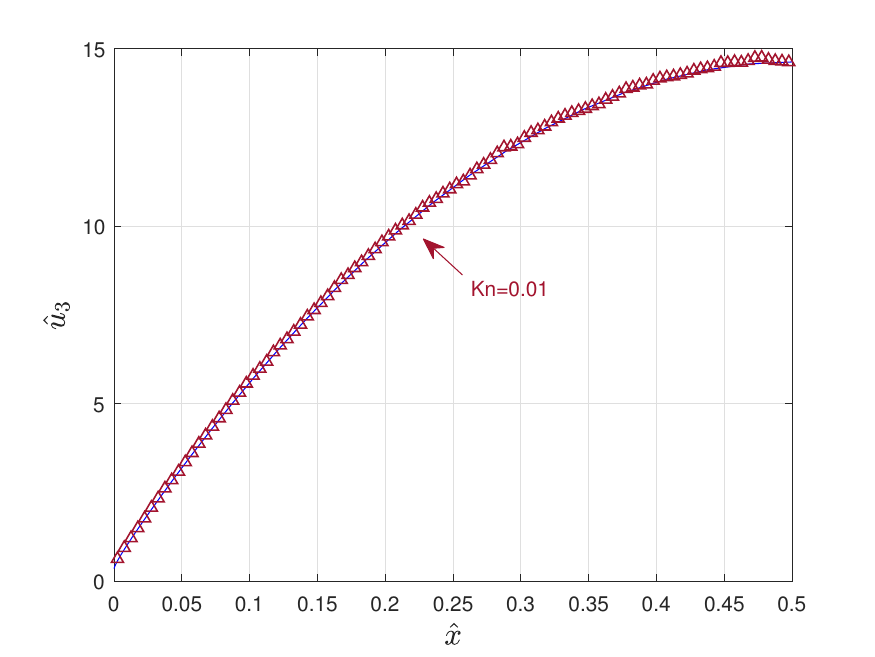}\\
	\includegraphics[width=0.45\textwidth]{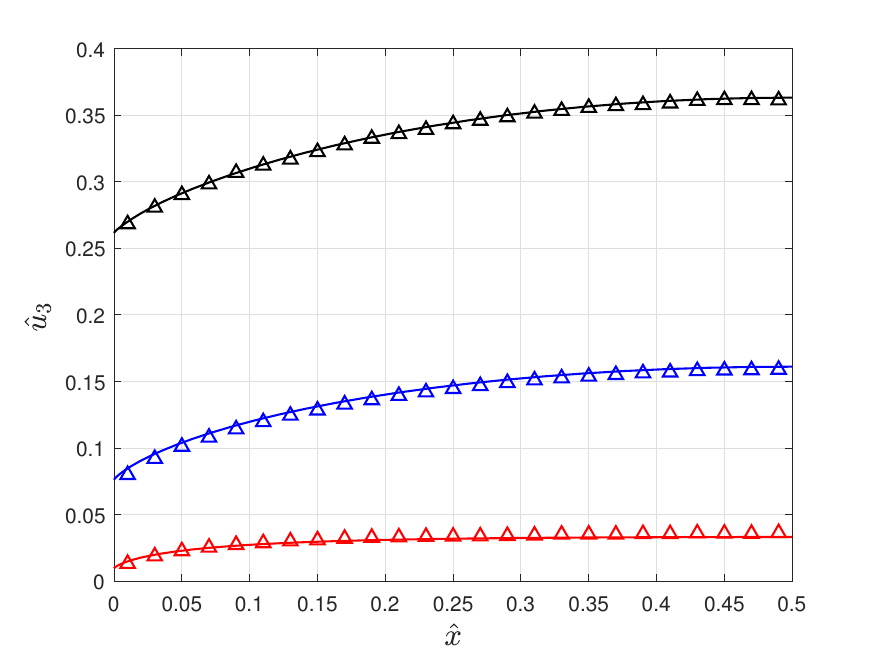}
  \includegraphics[width=0.45\textwidth]{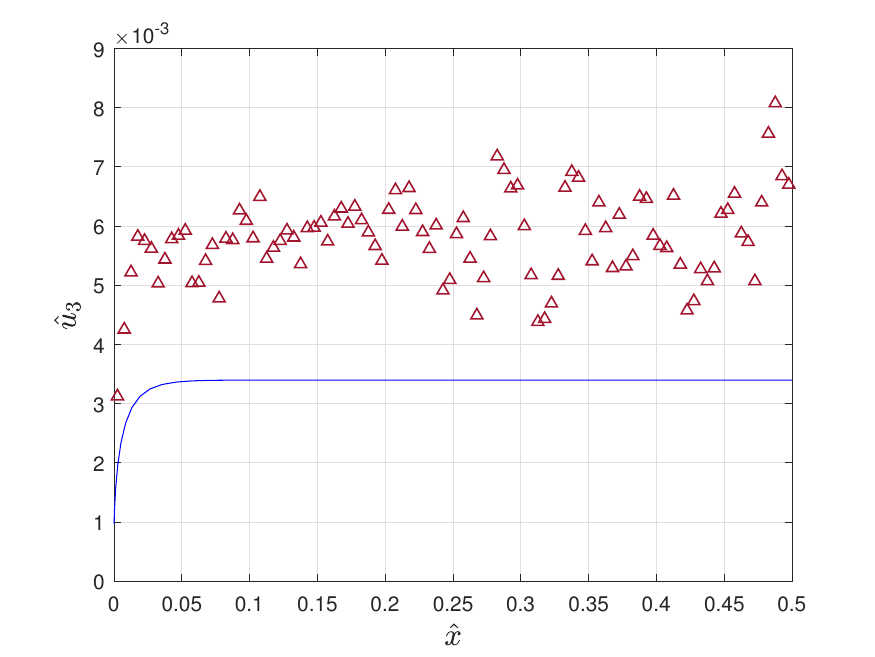}
	\caption{Velocity profiles in the 1D Poiseuille flow (top) and thermal transpiration (bottom) when $\text{Kn}=0.01, 0.1, 1, 10$. }
	\label{MP}
\end{figure}

Velocity profiles of the Poiseuille flow and thermal transpiration obtained from GSIS-LVDSMC and GSIS-DVM are compared in Fig.~\ref{MP}. In the Poiseuille flow, excellent agreements of velocity profiles are seen for all Knudsen numbers. In the thermal transpiration, although the high accuracy of GSIS-LVDSMC is demonstrated in the free-molecular and transition flow regimes, it predicts higher values of creep velocity than those from GSIS-DVM when $\text{Kn}=0.01$. This is because the flow velocity approaches zero when the Knudsen number decreases, and large fluctuation leads to inaccuracy of the macroscopic properties and high-order terms sampled from LVDSMC, when the sample size are not large enough.

Next, we calculate the mass flow rate of the Poiseuille flow to quantify the accuracy and efficiency of GSIS-LVDSMC, where the dimensionless mass flow rate $\dot{m}_p$ is defined as 
\begin{equation}
\dot{m}_p = 2\int \hat{u}_3d\hat{x}_1.
\end{equation}
Table~\ref{tab1} compares the essential simulation parameters and computational efficiency between GSIS-LVDSMC and LVDSMC. It is shown that GSIS-LVDSMC and LVDSMC have the same efficiency when the system is in high nonequilibrium, e.g., when $\text{Kn}=1$ and 10, these two methods require the same number of times steps to converge in both transition and steady states. Besides, since the macroscopic equations have to be solved at each iteration step in GSIS-LVDSMC, it takes a little more CPU time than LVDSMC. However, as the Knudsen number decreases, the computational cost increases dramatically in LVDSMC. Remarkably, the GSIS-LVDSMC improves the efficiency by nearly 50 times when $\text{Kn}=0.01$; meanwhile, the accuracy is also improved, leading to the relative error lower than 0.02\%, compared to 6.22\% in LVDSMC.

\begin{table*}[t]
	\footnotesize
	\caption{\label{tab1}Simulation parameters in the GSIS-LVDSMC and LVDSMC (values in parentheses)  used in the 1D Poiseuille flow.}
	\centering
	\begin{tabular}{|c|c|c|c|c|}
		\hline
							&   $\text{Slip flow}$   &   \multicolumn{3}{c|}{$\text{Transition flow}$} \\ \hline
		$\text{Kn}$         		   &   0.01  			&   0.1              &   1           	   &   10 \\ \hline
		Number of cells ($N$)        		   &   100 (300)        &   \multicolumn{3}{c|}{25 (25)}      \\ \hline
		$N\times\Delta \hat{t}$     &   0.05 (0.5)       &   \multicolumn{3}{c|}{1.0 (1.0)}     \\ \hline
		Number of time steps in transition state   &   \makecell[c]{$10^4$\\($3\times 10^5$)}   &   \makecell[c]{2500\\($2\times 10^4$)}   &  \makecell[c]{2500\\(5000)} & \makecell[c]{2500\\($2500$)}  \\ \hline
		Number of time steps in steady state        &   \makecell[c]{$7\times10^4$ \\ ($2\times10^5$)}  &   \makecell[c]{$2.5\times10^5$\\($3\times10^5$)}  &   \makecell[c]{$3\times10^5$\\($3\times10^5$)}   &   \makecell[c]{$3\times10^5$\\($3\times10^5$)} \\ \hline
		CPU Time$^a$    			   &   \makecell[c]{588 s\\(8 h)}     &   \makecell[c]{194 s\\(205 s)}    &   \makecell[c]{50 s\\(45 s)}   	   &   \makecell[c]{52 s\\(36s)} \\ \hline
		Number of particles$^b$   	   &   \makecell[c]{13000\\(14000)}    &  \makecell[c]{400\\(400)}   	 &   \makecell[c]{200\\(200)}   	   &   \makecell[c]{400\\(400)} \\ \hline
		Error relative to GSIS-DVM   		   &   \makecell[c]{0.019\%\\(6.22\%)} &   \makecell[c]{0.6\%\\(2\%)}   	 &   \makecell[c]{1\%\\(0.1\%)}  	   &   \makecell[c]{0.5\%\\(0.1\%)} \\ \hline
		\multicolumn{5}{@{}l}{\scriptsize$a.$ All simulations are done on a single core of an Intel(R) Core(TM) i7-10700K CPU @ 3.80GHz processor} \\
		\multicolumn{5}{@{}l}{\scriptsize$b.$ The time-averaged number of total deviational particles}
	\end{tabular}
\end{table*}

\subsubsection{2D cases}

Consider the 2D Poiseuille flow in an infinite long channel with a square cross section. Due to symmetry, only the lower left quarter of the cross-section is simulated ($0\le\hat{x}_1\le0.5$, $0\le\hat{x}_2\le0.5$), which is uniformly discretized into $N_{x_1}\times N_{x_2}$  distributed spatial cells. The bottom ($0\le\hat{x}_1\le0.5$, $\hat{x}_2=0$) and left boundaries ($\hat{x}_1=0$, $0\le\hat{x}_2\le0.5$) are fully diffuse, while the top ($0\le\hat{x}_1\le0.5$, $\hat{x}_2=0.5$) and right ($\hat{x}_1=0.5$, $0\le\hat{x}_2\le0.5$) boundaries satisfy the symmetric conditions.

\begin{figure}[t]
	\centering
	\subfloat[$\text{Kn}=10$]{\includegraphics[width=0.4\textwidth]{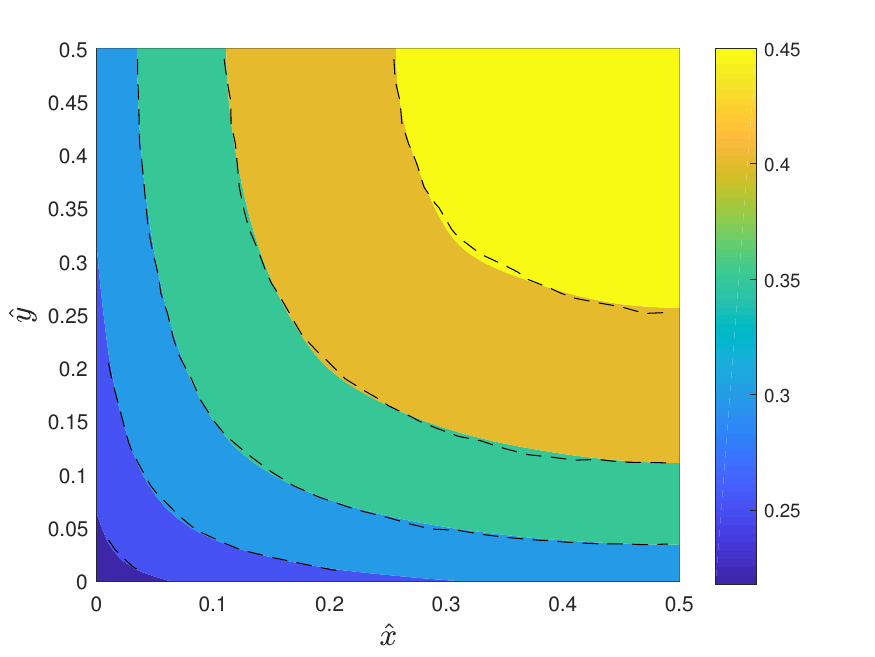}}
	\subfloat[$\text{Kn}=1$]{\includegraphics[width=0.4\textwidth]{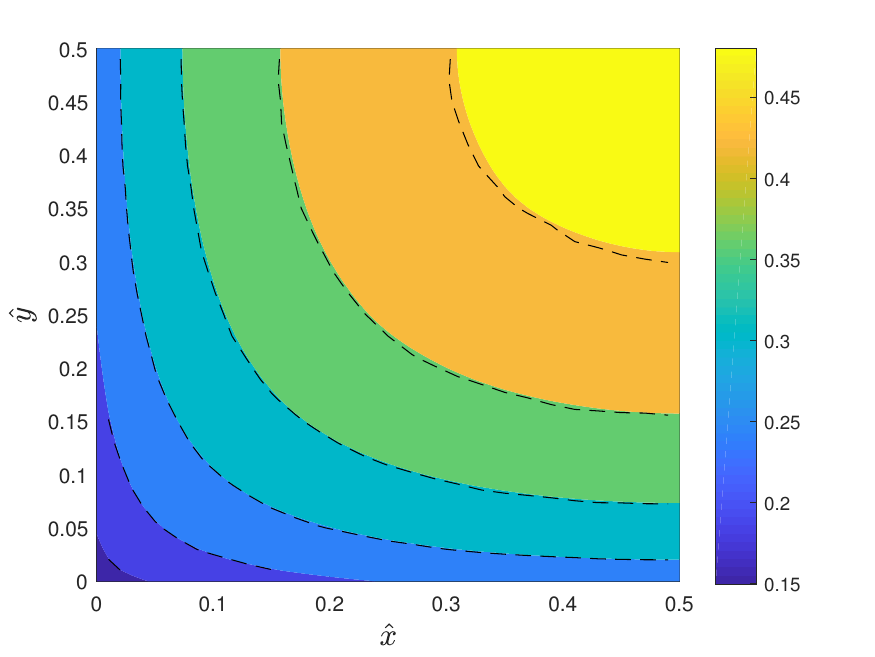}}\\
	\subfloat[$\text{Kn}=0.1$]{\includegraphics[width=0.4\textwidth]{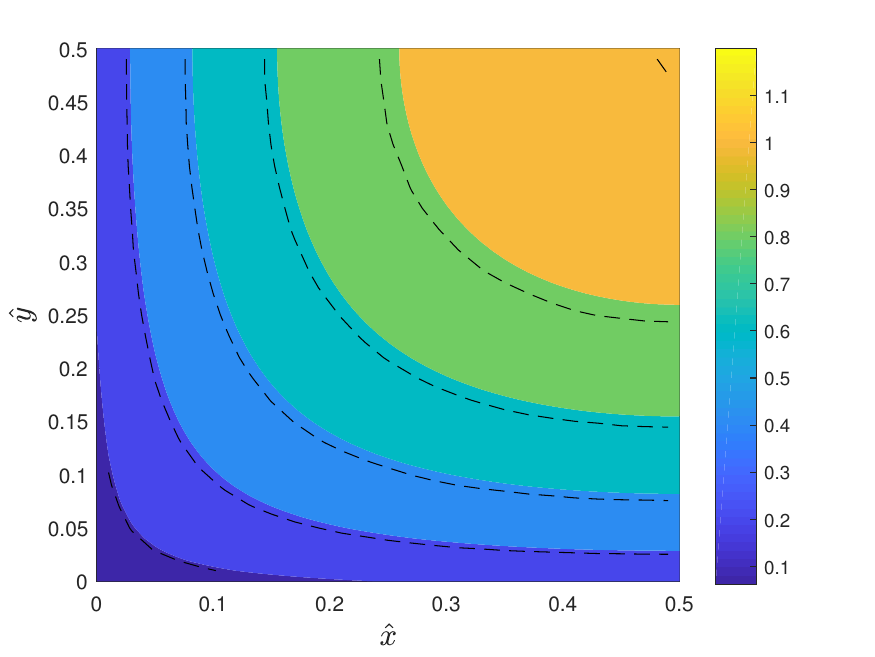}}
	\subfloat[$\text{Kn}=0.01$]{\includegraphics[width=0.4\textwidth]{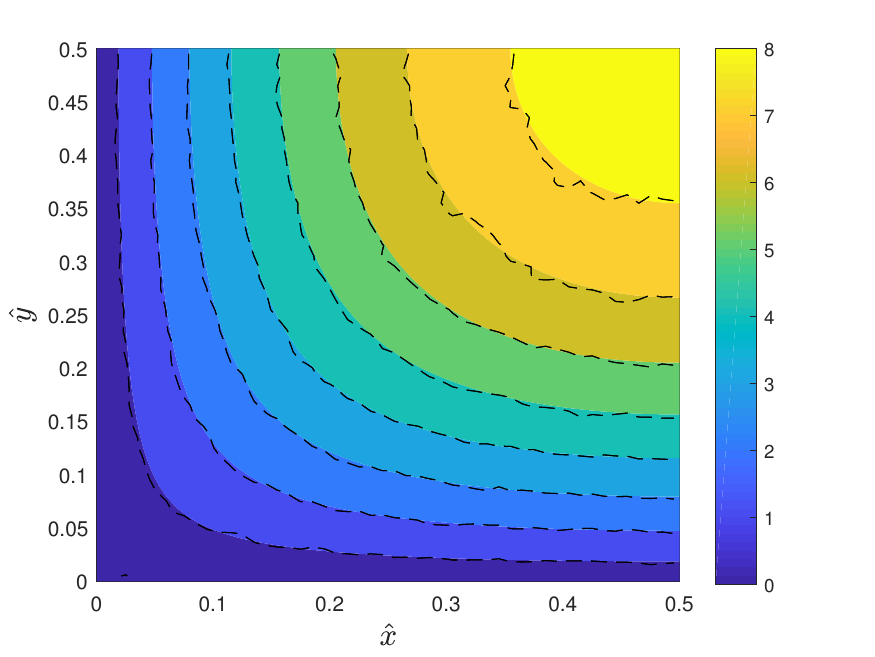}}
	\caption{2D velocity contour and isoline in the 2D Poiseuille flow. Contour maps: the reference solution from GSIS-DVM. Dotted contour lines: GSIS-LVDSMC results.
	}
	\label{2D_contour}
\end{figure}

Based on the central difference scheme,   Eq.~\eqref{eq:SIS} can be solved as:
\begin{equation}
\begin{aligned}[b]
&~2\left(\alpha+\frac{1}{\alpha}\right)\hat{U}_{i,j}-\alpha\hat{U}_{i,j+1}-\alpha\hat{U}_{i,j-1}-\frac{1}{\alpha}\hat{U}_{i+1,j}-\frac{1}{\alpha}\hat{U}_{i-1,j}\\
=&~\delta_{rp}\Delta x_1\Delta x_2+\alpha\left(M_{i,j+1}-2M_{i,j}+M_{i,j-1}\right)\\
&+\frac{1}{4}\left(R_{i+1,j+1}-R_{i-1,j+1}-R_{i+1,j-1}+R_{i-1,j-1}\right)\\
&+\frac{1}{\alpha}\left(N_{i+1,j}-2N_{i,j}+N_{i-1,j}\right), \quad i=2,...,N_{x_1}, \quad j=2,...,N_{x_2},
\end{aligned}
\label{eq:LR_SIS}
\end{equation}
where $\hat{U}$ represents $\hat{u}_3$ for simplicity, $M,R,N$ represent the high-order terms $\frac{1}{4}F_{2,0,1}$, $\frac{2}{4}F_{1,1,1}$, $\frac{1}{4}F_{0,2,1}$ in Eq.~\eqref{eq:HOT}, respectively; $i,~j$ are the index for the cells in $x_1$ and $x_2$ directions, respectively, and $\alpha=\Delta x_2/\Delta x_1$ is the cell aspect ratio.
The velocities sampled from LVDSMC and the symmetric conditions provide the boundary conditions:
\begin{itemize}
	\setlength{\itemsep}{0pt}
	\setlength{\parsep}{0pt}
	\setlength{\parskip}{0pt}
	\item When $i=1$ or $j=1$, the flow velocities $\hat{U}_{i,j}$ sampled at each time step and the time-averaged high-order terms $M,R,N$ are applied to solve the macroscopic equation~\eqref{eq:LR_SIS}.
	\item When $i=N_{x_1}$, the symmetric conditions read: $\hat{U}_{i+1,j}=\hat{U}_{i,j}$, $N_{i+1,j}=N_{i,j}$, $R_{i+1,j}=-R_{i,j}$; When $j=N_{x_2}$, the symmetric conditions read $\hat{U}_{i,j+1}=\hat{U}_{i,j}$, $M_{i,j+1}=M_{i,j}$, $R_{i,j+1}=-R_{i,j}$.
\end{itemize}
The relative error in flow velocity between two successive time steps defined in Eq.~\eqref{eq:relative_error_1D} is extended in 2D cases as:
\begin{equation}
\varepsilon = \iint\left|\frac{\hat{U}^{(k+1)}}{\hat{U}^{(k)}}-1\right| d\hat{x}_1d\hat{x}_2,
\label{eq:Relative_error_2D}
\end{equation}
where the velocity $\hat{U}$ is the time-averaged value, and the simulation is switched to steady state and terminated when $\varepsilon<10^{-3}$ and $\varepsilon<10^{-6}$, respectively. 

Figure \ref{2D_contour} shows the velocity contours from GSIS-LVDSMC and GSIS-DVM, when $\text{Kn}=0.01,0.1,1$, and 10. Good agreements are achieved for all Knudsen numbers. The system parameters, computational accuracy and efficiency are compared in Table~\ref{tab2}. Similar to the 1D case, the efficiency of GSIS-LVDSMC is approximately the same as that of LVDSMC when the Knudsen number is large. When $\text{Kn}=0.01$, results predicted by LVDSMC are not converged after $N_{\text{step,max}}=5\times10^5$ steps, which has around 42\% relative difference to the reference solution. On the contrary, GSIS-LVDSMC obtains accurate results within $4\times10^4$ time steps. Therefore, the GSIS-LVDSMC is at least 50 times faster than LVDSMC.

\begin{table*}[t]
	\footnotesize
	\caption{\label{tab2}Comparison between GSIS-LVDSMC and LVDSMC (values in parentheses) algorithms for the 2D Poiseuille flow in square channel.}
	\centering
	\begin{tabular}{|c|c|c|c|c|}
		\hline
							&   $\text{Slip flow regime}$   &   \multicolumn{3}{c|}{$\text{Transition flow regime}$} \\ \hline
		$\text{Kn}$         		   &   0.01  			&   0.1              &   1           	   &   10 \\ \hline
		Number of cells $(N_{x_1}\times N_{x_2})$        		   &   \makecell[c]{$50\times50$\\($100\times100$)}   &   \multicolumn{3}{c|}{\makecell[c]{$25\times25$\\($25\times25$)}}     \\ \hline
		 $N_{x_1}\times\Delta \hat{t}$     &   0.1 (1.0)        &   \multicolumn{3}{c|}{1.0 (1.0)}    \\ \hline
Number of time steps in transition state   &   \makecell[c]{$100$\\($4\times 10^4$)}   &   \makecell[c]{200\\($4000$)}   &  \makecell[c]{800\\(1000)} & \makecell[c]{1000\\($1000$)}  \\ \hline
		Number of time steps in steady state       &   \makecell[c]{$4\times10^4$\\($5\times10^5$)}  &   \makecell[c]{$9\times10^4$\\($10^5$)}  &   \makecell[c]{$1.3\times10^5$\\($1.3\times10^5$)}   &   \makecell[c]{$1.4\times10^5$\\ ($1.4\times10^5$)} \\ \hline
		CPU Time    			   &   \makecell[c]{2 h\\(108 h)}     &   \makecell[c]{1700 s\\(2000 s)}    &   \makecell[c]{1000 s\\(900 s)}   	   &   \makecell[c]{1200 s\\(1000s)} \\ \hline
		Number of particles   	   &   \makecell[c]{25000\\(150000)}    &   \makecell[c]{6000\\(6000)}   	 &   \makecell[c]{5000\\(5000)}   	   &   \makecell[c]{6000\\(6000)} \\ \hline
		Error relative to GSIS-DVM &   \makecell[c]{0.9\%\\(42\%)} &   \makecell[c]{3.26\%\\(1.24\%)}   	 &   \makecell[c]{0.45\%\\(0.05\%)}  	   &   \makecell[c]{0.14\%\\(0.03\%)} \\ \hline
	\end{tabular}
\end{table*}

\subsection{Convergence rate}

\begin{figure}[p]
	\centering
	\subfloat[]{\includegraphics[width=0.45\textwidth]{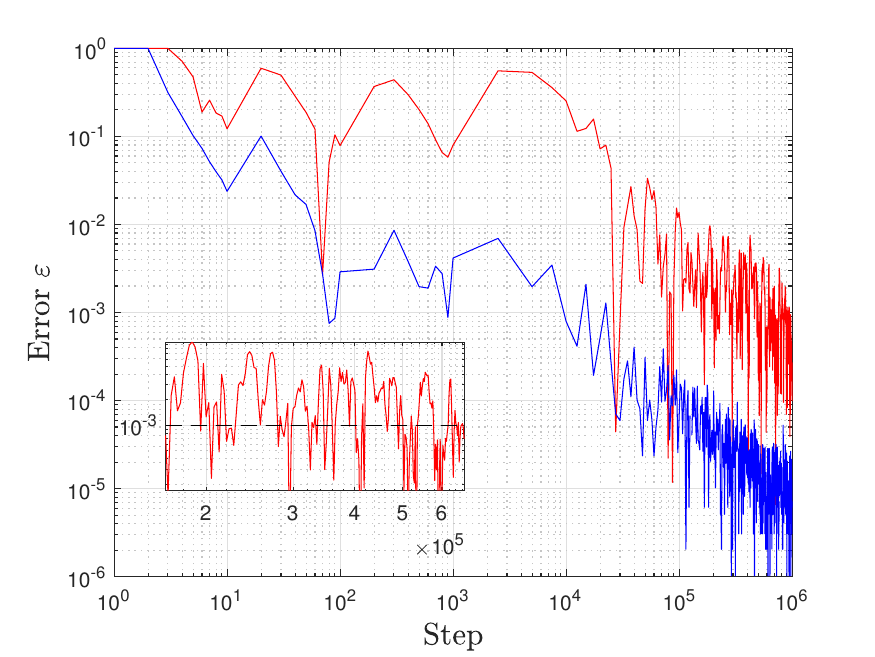}}
	\subfloat[]{\includegraphics[width=0.45\textwidth]{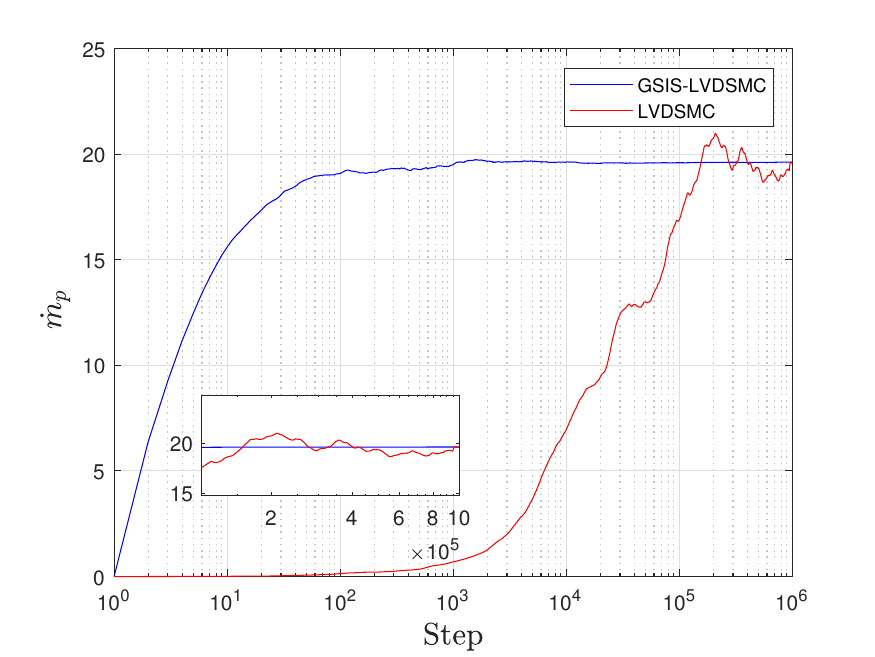}}
	\caption{History of convergence in the 1D Poiseuille flow solved by GSIS-LVDSMC (blue lines) and LVDSMC (red lines), when $\text{Kn}=0.01$. (a) relative error $\varepsilon$~\eqref{eq:Relative_error_2D} and (b)  mass flow rate. }
	\label{mpstep-convergence}
\end{figure}

\begin{figure}[p]
	\centering   
	\subfloat[$\text{Kn}=0.1$]{\includegraphics[width=0.45\textwidth]{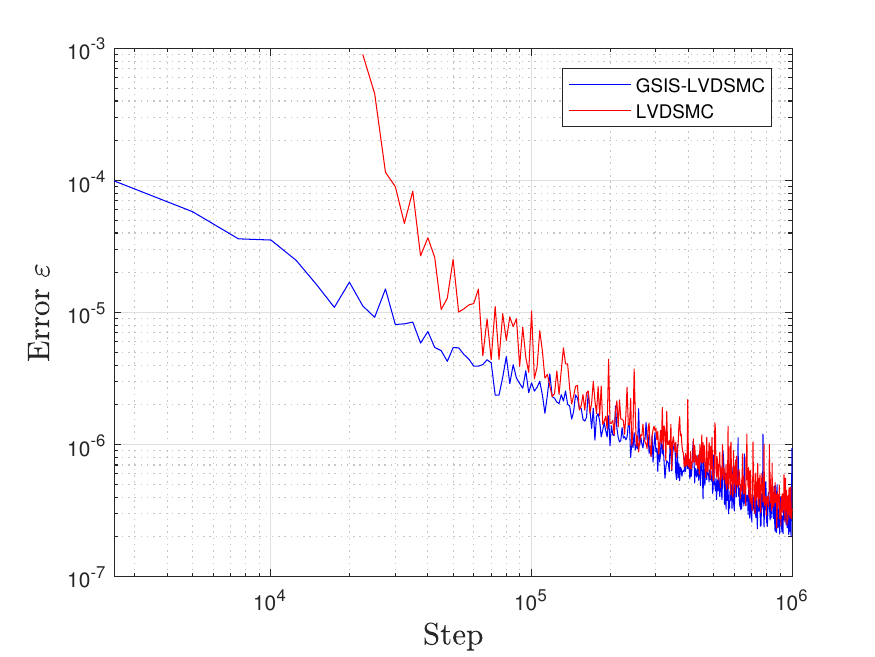}}
	\subfloat[$\text{Kn}=0.01$]{\includegraphics[width=0.45\textwidth]{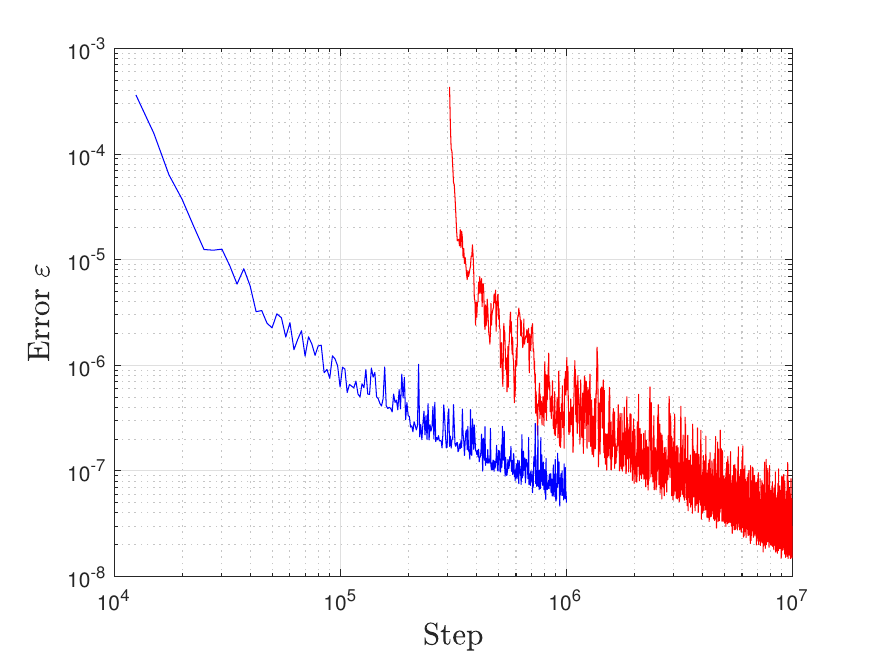}}\\
	\caption{The relative error $\varepsilon$ as a function of the iteration step, in the 1D Poiseuille flow solved by GSIS-LVDSMC (blue lines) and LVDSMC (red lines), when (a) $\text{Kn}=0.1$ and (b) $\text{Kn}=0.01$. The sampling starts when the steady-state is reached. }
	\label{convergence}
\end{figure}

We take the 1D Poiseuille flow to further discuss the mechanisms of convergence-boosting in GSIS-LVDSMC. As shown in Table \ref{tab1}, the numbers of iteration steps required in GSIS-LVDSMC and LVDSMC are approximately the same when $Kn$ is large, implying the same computational efficiency in simulating the high nonequilibrium gas flow. However, in the slip flow regime ($\text{Kn}=0.01$), GSIS-LVDSMC is about 50 times more efficient than LVDSMC. 

Figure \ref{mpstep-convergence} compares the convergence history of GSIS-LVDSMC and LVDSMC when $\text{Kn}=0.01$. Both the flow velocity and the mass flow rate are time-averaged values sampled from the beginning of  simulations. Therefore, the number of time steps for the system to reach the steady state can be determined. The GSIS-LVDSMC takes around $10^4$ iteration steps to make $\varepsilon$ below $10^{-3}$ and $\dot{m}_p$ converged. However, $3\times10^5$ time steps are required in LVDSMC to reach the same criteria, yet the oscillation in mass flow rate is still much more stronger. 
This can be understood as follows. In such a slip flow regime, frequent intermolecular collisions slow down the evolution of the gas flow from its initial state, i.e., slow down the information exchange across the computational domain. Also, LVDSMC introduces significant fluctuations in the flow velocities that lead to the oscillation in mass flow rate, as shown in Fig.~\ref{mpstep-convergence}(b). On the contrary, due to the coupling of macroscopic synthetic equations (which is of diffusion-type), which exchange the gas information across the whole computational domain, correct the gas properties, and drive the molecular distribution to the final solution quickly, the GSIS-LVDSMC greatly reduce the simulation step.

Figure~\ref{convergence} compares the decay of relative error $\varepsilon$ sampled from the two algorithms, when the steady state is reached. When $\text{Kn}=0.1$, the GSIS-LVDSMC and LVDSMC need about $2.5\times10^5$ and $3\times10^5$ time steps, respectively, to make $\varepsilon<10^{-6}$. When $\text{Kn}=0.01$, the time steps become  $7\times10^4$ and $2\times10^5$, respectively. Theoretically, there are two main factors that could influence the number of time steps: the particle number in each cell, and the correlation between two successive sampling steps. It can be seen from the Table~\ref{tab1} that, the  particle number are almost the same in GSIS-LVDSMC and LVDSMC. Therefore, we draw the conclusion that the coupling of synthetic equations and adjusting the simulation particles accordingly reduce the correlation of successive time steps and hence the fluctuations. Clearly, GSIS-LVDSMC not only accelerates the transition state of the particle method, but also achieves higher accuracy with fewer sample steps in steady state.

\subsection{Cell size and time step}

In LVDSMC, due to the splitting of advection and collision, the cell size and time step should be smaller than the kinetic scales (i.e., the mean free path and mean collision time of gas molecules, respectively). Consequently, the cell number and the computational cost increase dramatically when the system is in the near continuum regime. On the contrary, in GSIS-LVDSMC, these restrictions are removed by coupling the synthetic equations, which the Navier-Stokes equations are predominant in the continuum limit so that the hydrodynamic scale (which is much larger than the kinetic scale) can be used. Detailed evidence are given below.

\begin{figure}[t]
	\centering
	\subfloat[]{\includegraphics[width=0.45\textwidth]{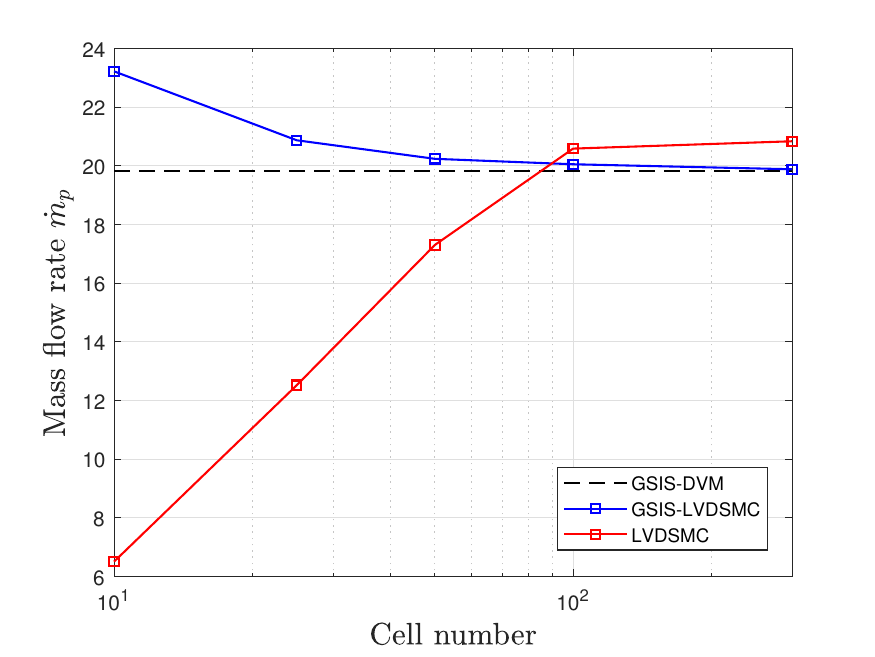}}
	\subfloat[]{\includegraphics[width=0.45\textwidth]{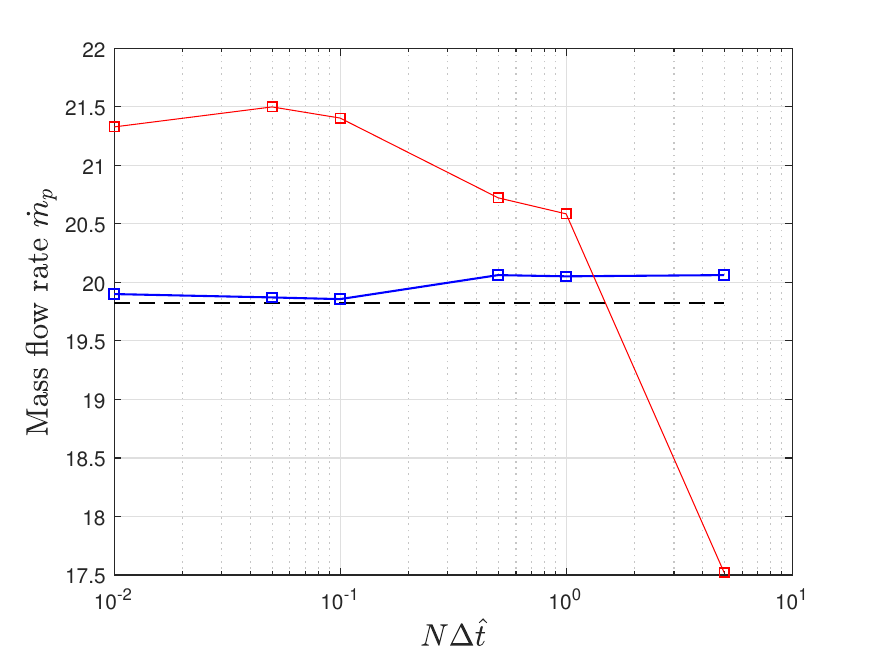}}
	\caption{The mass flow rate for 1D Poiseuille flow obtained by GSIS-LVDSMC (blue lines) and LVDSMC (red lines) changes with (a) the cell number and (b) the time step, when $\text{Kn}=0.01$. }
	\label{GRFMFR}
\end{figure}

Figure \ref{GRFMFR}(a) shows the mass flow rate obtained from the two methods, when the cell number is changed from 10 to 300, which correspond to the cell size 10 and 1/3 times of the molecular mean free path, respectively. It can be seen that the mass flow rates from LVDSMC with cell size larger than the mean free path are wrong, and even when the cell size decreases to 1/3 of the mean free path the result has 5\% relative difference to the reference solution. On the other hand, results from GSIS-LVDSMC converge much faster and monotonically when the cell size is decreased. The relative errors in mass flow rate are less than 1\% as long as the cell size is smaller than 2 times of the mean free path in this problem.

Figure~\ref{GRFMFR}(b) shows the mass flow rate of 1D Poiseuille flow with the time step $N\Delta\hat{t}$ varying from $0.01$ to $5$, where $N$ is the number of grid cell ($N=100$ and $N=300$ are used in GSIS-LVDSMC and LVDSMC, respectively). Therefore, $N\Delta\hat{t}=1$ means that a particle with the most probable speed $c_0$ travels a distance of one simulation cell during the time step $\Delta\hat{t}$. In GSIS-LVDSMC, results are not sensitive to the time step, and the relative error in mass flow rate is around 1\% even when $\Delta\hat{t}$ is 5 times of the mean collision time. However, as it is commonly acknowledged, the time step in LVDSMC should be smaller than 1/3 of the mean collision time to guarantee a reliable result, which corresponds to $N\Delta\hat{t}=1$ here (grid size is 1/3 of the mean free path in this case). As shown in Fig.~\ref{GRFMFR}(b), larger $\Delta\hat{t}$ in LVDSMC leads to wrong mass flow rate. However, smaller $\Delta\hat{t}$ requires so tremendous computational cost that the results of LVDSMC are not converged within the maximum simulation time step $N_{\text{step,max}}$.


\section{Numerical tests for planar Fourier flow}\label{sec_fourier}

Consider the planar Fourier flow of gas between two infinite parallel plates located at $\hat{x}_1=0$ and $\hat{x}_1=1$ with perturbed temperature $\tau=-0.5$ and 0.5 respectively. Assuming the symmetric condition $\hat{h}(-\hat{c}_1,\hat{c}_2,\hat{c}_3)=-\hat{h}(\hat{c}_1,\hat{c}_2,\hat{c}_3)$ can be applied at $\hat{x}_1=0.5$. Based on the synthetic equations Eqs.~\eqref{eq:Evolution}, \eqref{eq:stress} and \eqref{eq:heatflux}, we have 
$\hat{\bm{u}}=0$, $\hat{\sigma}_{ij}=0~(i\neq j)$, $\hat{q}_2=\hat{q}_3=0$, ${\partial\hat{q}_1}/{\partial\hat{x}_1}=0$, and the constitutive relations are simplified to:
\begin{equation}
	\begin{aligned}[b]
		\frac{\partial}{\partial \hat{x}_1}\underbrace{\int2\left(\hat{c}_1^2-\frac{\hat{\bm{c}}^2}{3}\right)\hat{c}_1\hat{h}d^3\hat{c}}_{F_{\sigma_{11}}}&=-\delta_{rp}\hat{\sigma}_{11}, \\
		\frac{\partial}{\partial\hat{x}_1}\underbrace{\int\left[\left(\hat{c}_1^2-\frac{5}{6}\right)\left(\hat{\bm{c}}^2-\frac{3}{2}\right)-\hat{c}_1^2\right]\hat{h}d^3\hat{c}}_{F_{q_{1}}}+\frac{5}{4}\frac{\partial\tau}{\partial\hat{x}_1}&=-\delta_{rp}\bar{q}_1,
	\end{aligned}
	\label{eq:temperature_variation}
\end{equation}
where the high-order terms $F_{q_1}$ and $F_{\sigma_{11}}$ are statistically sampled from LVDSMC as follows:
\begin{equation}
	\begin{aligned}[b]
	F_{q_1}=&\frac{N_{\text{eff}}}{V_{\text{cell}}}\sum_{p\in cell}s_p\left[\left(\hat{c}_1^2-\frac{5}{6}\right)\left(\hat{{c}}^2-\frac{3}{2}\right)-\hat{c}_1^2\right],\\
	F_{\sigma_{11}}=&\frac{N_{\text{eff}}}{V_{\text{cell}}}\sum_{p\in cell}2s_p\left(\hat{c}_1^2-\frac{\hat{{c}}^2}{3}\right)\hat{c}_1.
	\end{aligned}
	\label{eq:Iq1}
\end{equation}

The problem can be solved either following the coupling algorithm strictly, or using the simplified procedure as follows. Since the heat flux $\hat{q}_1$ is a constant through the simulation domain, it is sampled according to Eqs. \eqref{eq:heatflux} and \eqref{eq:fd} as:
\begin{equation}
	\begin{aligned}
		\hat{q}_1&=-\frac{1}{\delta_{rp}}\bigg\langle\frac{\partial}{\partial \hat{x}_1}\frac{N_{\text{eff}}}{V_{\text{cell}}}\sum_{p\in \text{cell}} s_p\left(\hat{\bm{c}}^2-\frac{5}{2}\right)\hat{c}_1^2\bigg\rangle,
	\end{aligned}
\label{eq:constant_qx}
\end{equation}
where $\langle \cdot \rangle$ represents the ensemble average over the entire simulation domain. Therefore, both  $F_{q_{1}}$ and  $\hat{q}_1$ are sampled from the distribution of particles before solving macroscopic equations, which is indicated by step $(k+1/2)$ below. And then, the perturbed temperature at $(k+1)$-th step can be determined from Eq. \eqref{eq:temperature_variation} satisfying:
\begin{equation}
\tau^{(k+1)}=-\frac{4}{5}\delta_{rp}\bar{q}_1^{(k+1/2)}\left(\hat{x}_1-\frac{1}{2}\right)-\frac{4}{5}F_{q_{1}}^{(k+1/2)}.
\label{eq:numerical_solution}
\end{equation}
Meanwhile, the stress $\hat{\sigma}_{11}^{(k+1/2)}$ is obtained by calculating the spatial derivative of term $F_{\sigma_{11}}$ from Eq. \eqref{eq:temperature_variation}. Then, the perturbed density $\rho$ at $(k+1)$-th step  can be solved based on Eqs. \eqref{eq:Evolution} and \eqref{eq:gai}:
\begin{equation}
\rho^{(k+1)}+\tau^{(k+1)}+\hat{\sigma}_{11}^{(k+1/2)}=\text{const}=2\int\hat{c}_1^2\hat{h}d^3\hat{c},
\label{eq:sum_wtsgm}
\end{equation}
where the constant can be evaluated at $\hat{x}_1=0.5$ using the symmetric condition,  which is found to be zero.

The relative error between successive time steps $k$ and $k+1$ is defined as:
\begin{equation}
\varepsilon=\mathrm{max}\left\{\int\left|\frac{\rho^{(k+1)}}{\rho^{(k)}}-1\right|d\hat{x}_1,\ \int\left|\frac{\tau^{(k+1)}}{\tau^{(k)}}-1\right|d\hat{x}_1,\ \int\left|\frac{\hat{q}_1^{(k+1)}}{\hat{q}_1^{(k)}}-1\right|d\hat{x}_1\right\}.
\label{eq:residuals_Fourier}
\end{equation}
The simulation is switched to steady state and terminated when $\varepsilon<10^{-3}$ and $\varepsilon<10^{-6}$ are satisfied, respectively.

\subsection{Transition flow regime: $\text{Kn}=0.1, 1, 10$}

\begin{figure}[t]
	\centering
	\subfloat[]{\includegraphics[width=0.45\textwidth]{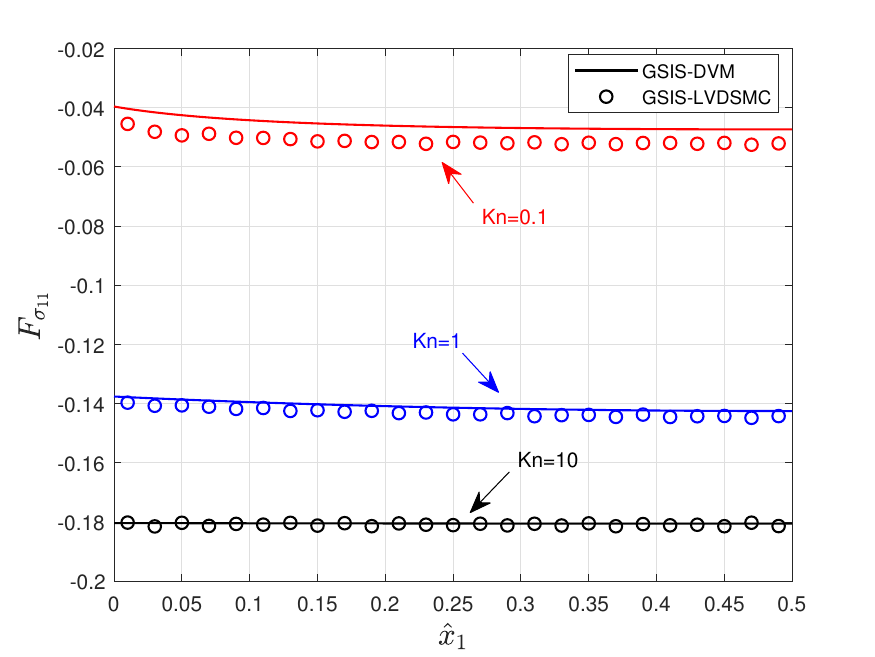}}
	\subfloat[]{\includegraphics[width=0.45\textwidth]{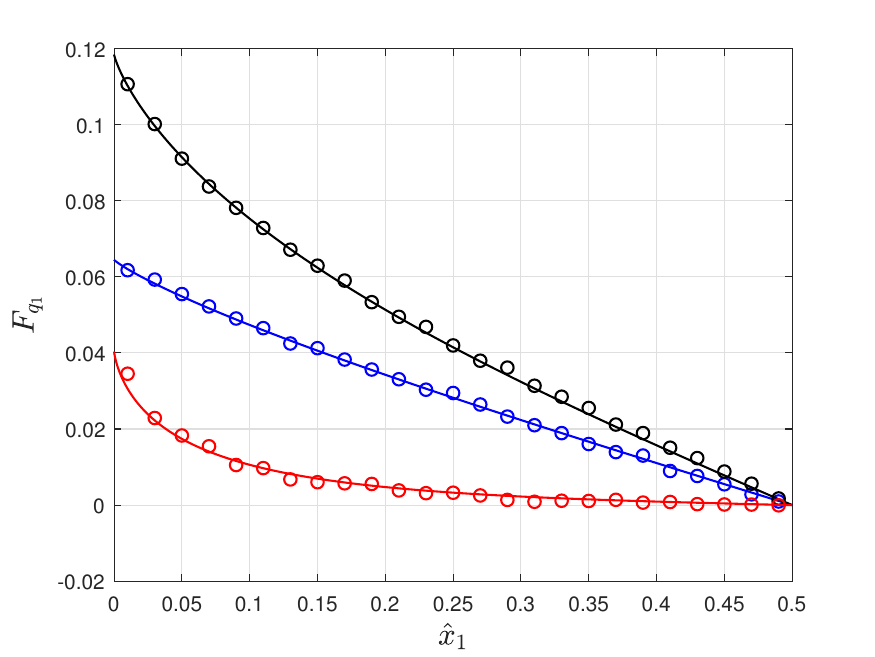}}\\
	\subfloat[]{\includegraphics[width=0.45\textwidth]{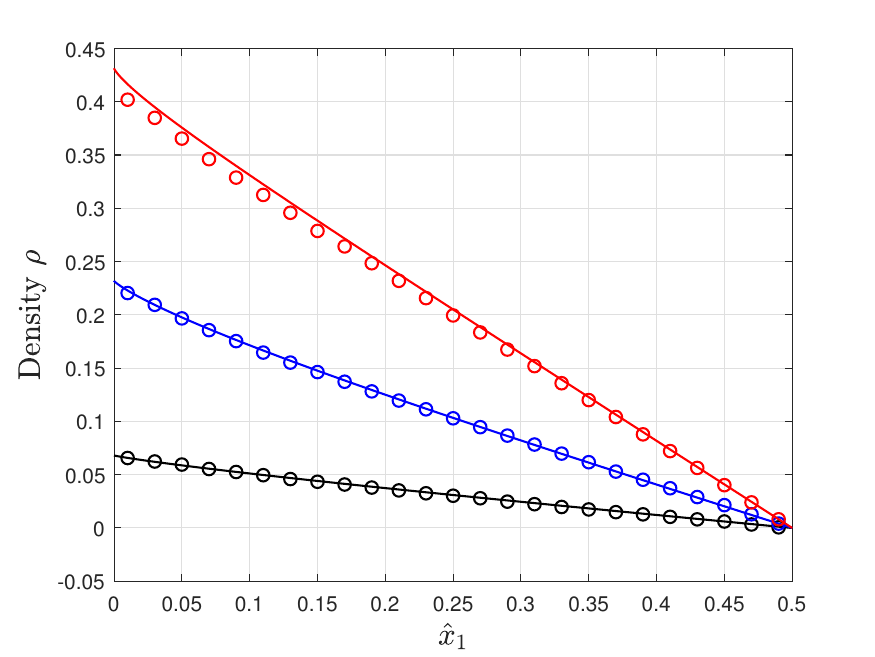}}
	\subfloat[]{\includegraphics[width=0.45\textwidth]{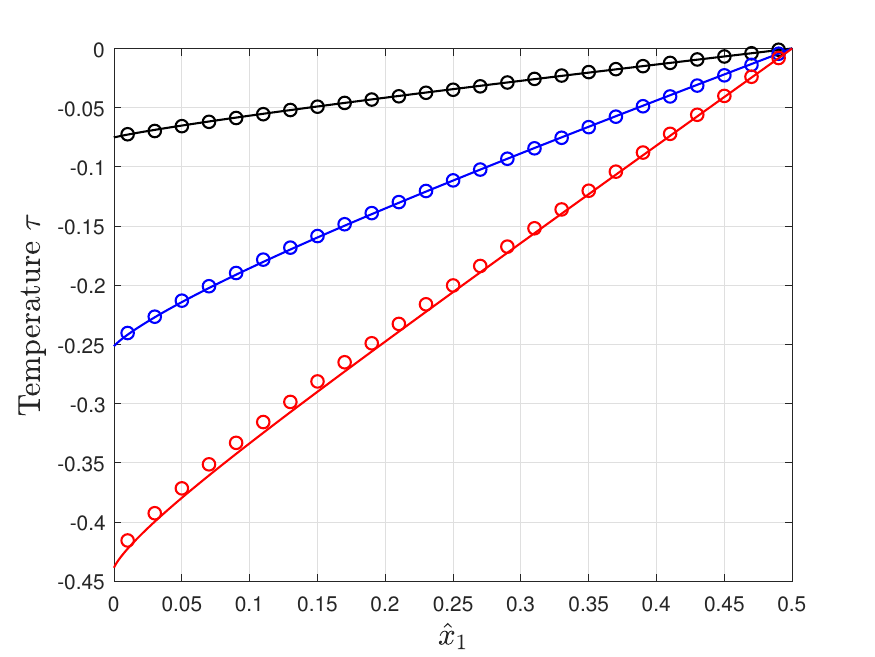}}
	\caption{The time-averaged properties obtained from GSIS-LVDSMC (lines) are compared to the reference solution from GSIS-DVM (circles). (a) $F_{\sigma_{11}}$ associated with the high-order term of stress, (b) $F_{q_1}$ associated with the high-order term of heat flux, (c) the perturbed density and (d) the perturbed temperature of planar Fourier flow. }
	\label{Fourier_transition}
\end{figure}

When $\text{Kn}=0.1, 1, 10$, the system parameters are set to be the same for both GSIS-LVDSMC and LVDSMC, where 25 uniform cells are used in the computational domain ($0\leq \hat{x}_1 \leq 0.5$) and $N\Delta\hat{t}=1$. Results obtained by GSIS-DVM are regarded as the reference solutions to assess the accuracy of GSIS-LVDSMC and LVDSMC. First, variations of terms $F_{\sigma_{11}}$ and $F_{q_1}$ are shown in Fig.~\ref{Fourier_transition}. Based on Eq.~\eqref{eq:temperature_variation}, high-order terms are the spatial derivatives of $F_{\sigma_{11}}$ and $F_{q_1}$, thus the nonequilibrium effects mainly occur through the constitutive relation of heat flux in the region several mean free path away from the walls. When the Knudsen number decreases, high-order terms begin to vanish in the central region of the Fourier flow. Figure~\ref{Fourier_transition}(c) and (d) show the perturbed converged density and temperature, where the accuracy of GSIS-LVDSMC is clearly demonstrated. Additionally, when compared to the reference solution obtained by GSIS-DVM, the heat flux predicted by GSIS-LVDSMC has only $0.92\%$, $0.27\%$ and $0.087\%$ relative difference, when $\text{Kn}=0.1,\,1$ and 10, respectively, which shows nearly the same accuracy compared with the original LVDSMC algorithm.

\begin{table*}[t]
	   \footnotesize
		\caption{\label{tab3}Parameters adopted in GSIS-LVDSMC and LVDSMC (values in parentheses) algorithms in the simulation of 1D Fourier flow.}
		\centering
	\begin{tabular}{|c|c|c|c|c|}
		\hline
	&   $\text{Slip flow}$   &   \multicolumn{3}{c|}{$\text{Transition flow}$} \\ \hline
		$\text{Kn}$         		   &   0.01  			&   0.1              &   1           	   &   10 \\ \hline
		Number of cells ($N$)        		   &   50 (300)        &   \multicolumn{3}{c|}{50(50)}      \\ \hline
		$N\times\Delta \hat{t}$     &   0.5 (1.0)       &   \multicolumn{3}{c|}{1.0 (1.0)}     \\ \hline
		Number of time steps in transition state   &   \makecell[c]{$1.25\times10^4$\\($3\times 10^5$)}   &   \makecell[c]{2000\\($3\times 10^4$)}   &  \makecell[c]{2000\\(4000)} & \makecell[c]{2000\\($2000$)}  \\ \hline
		Number of time steps in steady state        &   \makecell[c]{$7\times10^5$ \\ ($10^7$)}  &   \makecell[c]{$10^6$\\($3\times10^6$)}  &   \makecell[c]{$10^6$\\($10^6$)}   &   \makecell[c]{$10^6$\\($10^6$)} \\ \hline
		CPU Time    			   &   \makecell[c]{82 s\\(40 h)}     &   \makecell[c]{240 s\\(1536 s)}    &   \makecell[c]{97 s\\(81 s)}   	   &   \makecell[c]{201 s\\(76 s)} \\ \hline
		Number of particles   	   &   300 (13000)    &   160 (500)   	 &   200 (330)   	   &   600 (500) \\ \hline
		Error relative to GSIS-DVM   		   &   \makecell[c]{2.85\%\\(8.81\%)} &   \makecell[c]{0.92\%\\(0.32\%)}   	 &   \makecell[c]{0.27\%\\(0.34\%)}  	   &   \makecell[c]{0.087\%\\(0.032\%)} \\ \hline
	\end{tabular}
\end{table*}

The simulation parameters, especially the numbers of iteration step and computational cost, are summarized in Table~\ref{tab3}. Similar to the situations in the Poiseuille flow and thermal transpiration, the efficiency of GSIS-LVDSMC is close to that of LVDSMC when the Knudsen number is large, although the CPU time cost in GSIS-LVDSMC is relatively large due to the solving of synthetic equations. However, when $\text{Kn}=0.1$, the time steps ($3\times10^4$) in LVDSMC significantly increases, but the GSIS-LVDSMC takes only 2000 iteration steps to reach the steady state in the transition regime. Eventually, the CPU time can be reduced by 6 times when GSIS-LVDSMC is applied. The computational cost is reduced by about 1800 times when $Kn=0.01$, which will be even bigger when the Knudsen number is further reduced.

\subsection{Slip flow regime: $\text{Kn}=0.01$}

\begin{figure}[t]
	\centering
	\subfloat[$\text{Kn}=0.1$]{\includegraphics[width=0.45\textwidth]{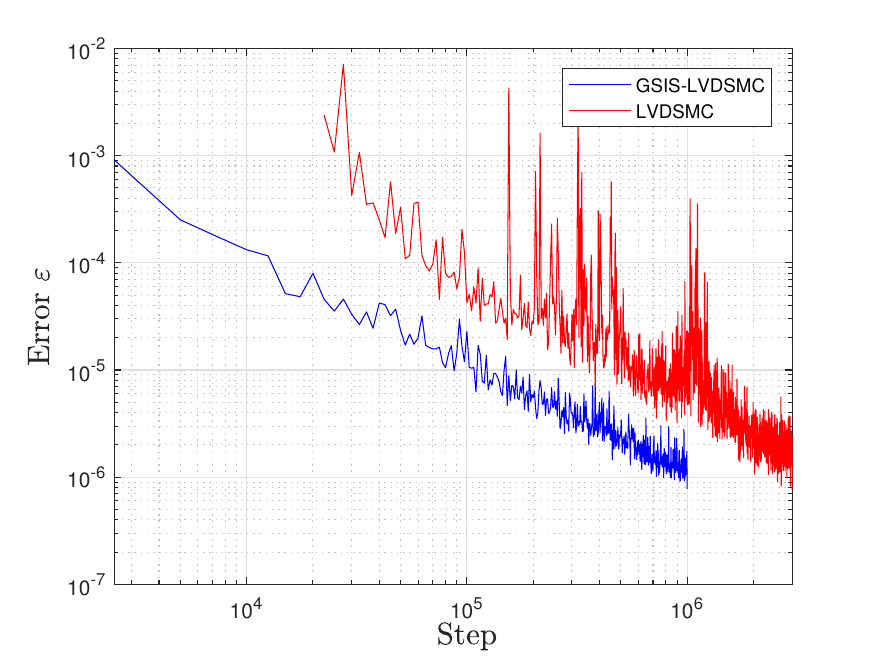}}
	\subfloat[$\text{Kn}=0.01$]{\includegraphics[width=0.45\textwidth]{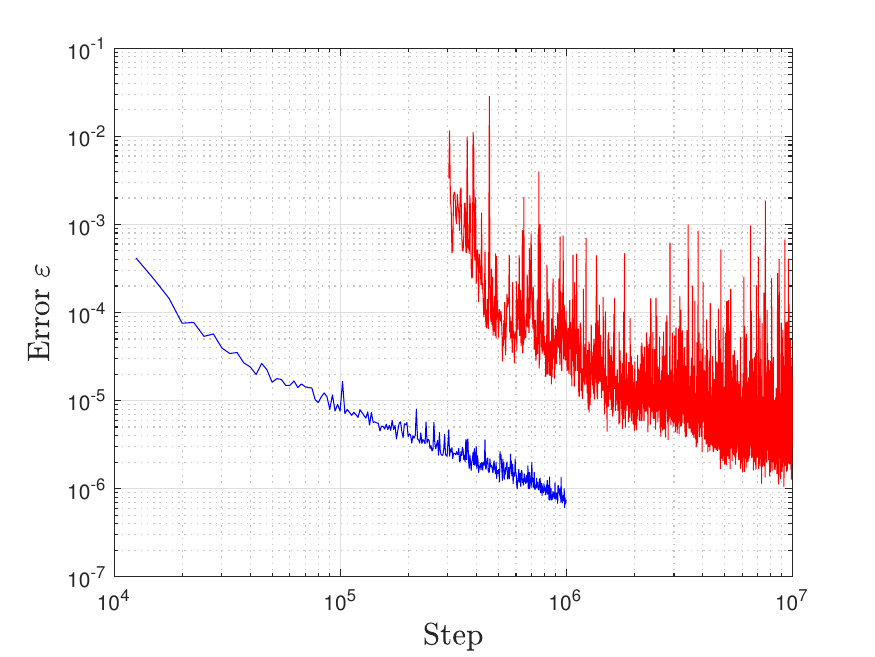}}
	\caption{Decay of the relative error $\varepsilon$ in steady state with the iteration step for planar Fourier flow obtained by GSIS-LVDSMC (blue lines) and LVDSMC (red lines). }
	\label{Fourier_residuals}
\end{figure}

\begin{figure}[h]
	\centering
{\includegraphics[width=0.45\textwidth]{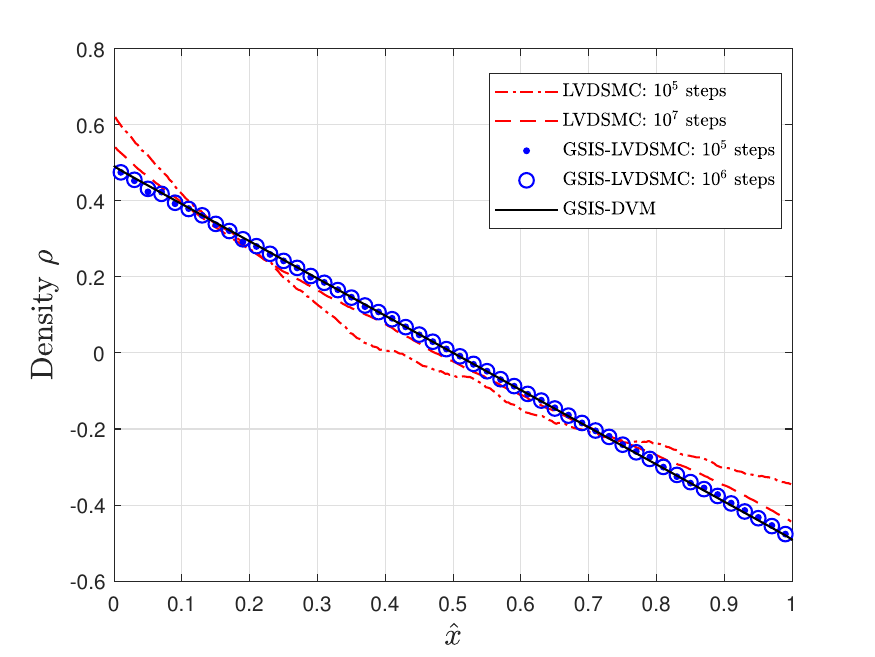}}
{\includegraphics[width=0.45\textwidth]{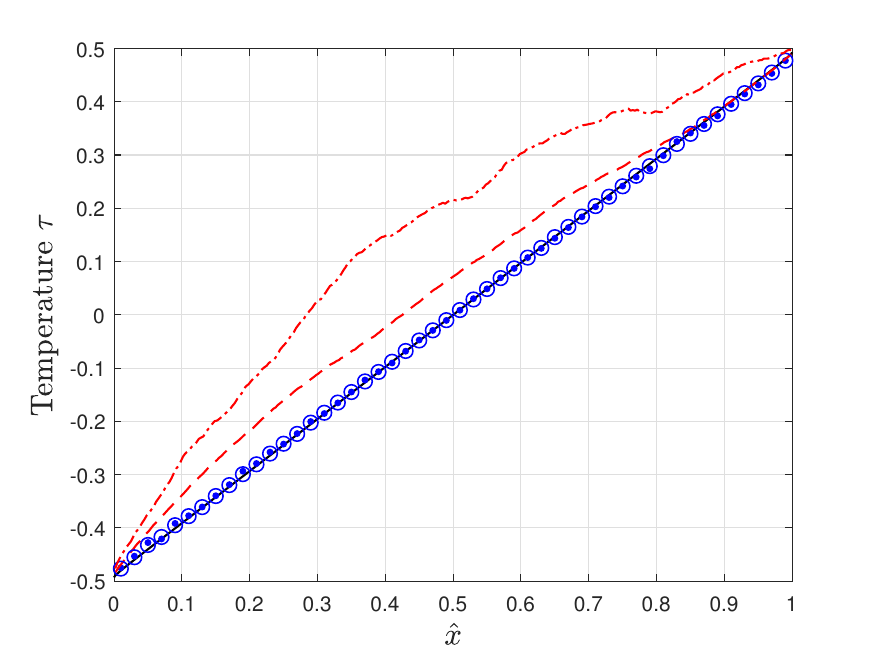}}
	\caption{Density and temperature profiles in the planar Fourier flow when $\text{Kn}=0.01$. Solid black lines: GSIS-DVM.  Red lines: LVDSMC after $10^6$ (dot dash line) and $10^7$ (dotted line) time steps. Dots and circles: GSIS-LVDSMC after $10^5$ and $10^6$ time steps, respectively. }
	\label{Fourier_results_Kn001}
\end{figure}

When $\text{Kn}=0.01$, due to the restriction on cell size and time step, 300 spatial cells and the time step $N\Delta\hat{t}=1$ are used in LVDSMC. In the GSIS-LVDSMC algorithm, 50 cells and $N\Delta\hat{t}=0.5$ are applied instead. Figure~\ref{Fourier_residuals} illustrates the difference in the decay of relative error $\varepsilon$. The discrepancy in  convergence history between the two methods appears when $\text{Kn}=0.1$, where GSIS-LVDSMC converges faster than LVDSMC by around $10^6$ iteration steps. When $\text{Kn}=0.01$, LVDSMC needs about 10 times of times step than GSIS-LVDSMC to reach the same relative error.

Besides, the fluctuation of LVDSMC is much more significant, although the number of simulation particles in each cell is 7 times than that in GSIS-LVDSMC. Figure~\ref{Fourier_results_Kn001} shows the perturbed density and temperature at different time steps, when $\text{Kn}=0.01$. GSIS-LVDSMC provides accurate solutions, with little fluctuations, after $10^5$ time steps, which have less than 3\% difference in heat flux compared to the reference solution. However, both the density and temperature from LVDSMC have significant fluctuations and discrepancy to the reference solutions, even $N_{\text{step,max}}=10^7$ and 40 hours CPU time have been spent. In the slip flow regime, LVDSMC takes large number of times step to make the influence from the walls pass through the entire bulk region, while the synthetic equations in GSIS-LVDSMC, because of its diffusion-type, help to pass the perturbation through the whole simulation domain immediately, thus boosting the convergence. Furthermore, the correction of macroscopic properties of the flow field also reduce fluctuations of the stochastic method, leading to a smaller sample size required in doing time average sampling.

\section{Conclusions}\label{sec_conclusion}

In summary, we have developed a GSIS to improve the computational efficiency of the LVDSMC method. The accuracy of the GSIS-LVDSMC coupling is validated in the Poiseuille flow, thermal transpiration, and Fourier flow. The fast convergence of GSIS-LVDSMC is achieved by solving the synthetic macroscopic equations at each time step, which not only explicitly contain the exact constitutive relations for shear stress and heat flux extracted from LVDSMC, but also asymptotically preserve the Navier-Stokes limit. The former feature guarantees the accuracy of the algorithm in all flow regimes, since the rarefaction effects are captured by  high-order terms of the constitutive relations extracted from LVDSMC, the latter removes the constraints that the spatial cell size should be smaller than the molecular mean free path, and the time step has to be smaller than mean collision time. With the coupling of synthetic equations, the velocity distribution of particles is adjusted according to the solutions and thus approaches to the steady state quickly. Therefore, the number of time step required in the transition researchgatestate is significantly reduced when the Knudsen number is small. Meanwhile, the sampling fluctuations are found to be much smaller in GSIS-LVDSMC in the near-continuum regime, thus its efficiency is further improved.

The proposed GSIS-LVDSMC coupling algorithm provides a framework to improve the computational efficiency of the conventional DSMC method in the near-continuum regime, which will have a much wider application scope than the low-variance version. We expect the essential idea in this scheme can be extended to efficiently and accurately solve the DSMC for multiscale hypersonic flows with chemical reaction, which has strong applications in space exploration and Mars' landing.

\section*{Acknowledgements}

This work is supported by the National Natural Science Foundation of China under the grant No.~12172162.

\section*{Declaration of interests}
The authors report no conflict of interest.


\bibliographystyle{IEEEtr}
\bibliography{ref}

\begin{thebibliography}{10}
\expandafter\ifx\csname url\endcsname\relax
  \def\url#1{\texttt{#1}}\fi
\expandafter\ifx\csname urlprefix\endcsname\relax\def\urlprefix{URL }\fi
\expandafter\ifx\csname href\endcsname\relax
  \def\href#1#2{#2} \def\path#1{#1}\fi

\bibitem{cercignani-2000}
C.~Cercignani, {Rarefied Gas Dynamics: From Basic Concepts to Actual
  Calculations (Cambridge Texts in Applied Mathematics, Series Number 21)}, 1st
  Edition, Cambridge University Press, 2000.

\bibitem{bird-1970}
G.~A. Bird, {Direct Simulation and the Boltzmann Equation}, Physics of Fluids
  13~(11) (1970) 2676.

\bibitem{wagner-1992}
W.~Wagner, {A convergence proof for Bird's direct simulation Monte Carlo method
  for the Boltzmann equation}, Journal of Statistical Physics 66~(3-4) (1992)
  1011--1044.

\bibitem{li-2013}
Z.~Li, M.~Fang, X.~Jiang, J.~Wu, {Convergence proof of the DSMC method and the
  Gas-Kinetic Unified Algorithm for the Boltzmann equation}, Science China
  Physics, Mechanics and Astronomy 56~(2) (2013) 404--417.

\bibitem{bird-1994}
G.~Bird, {Molecular Gas Dynamics and the Direct Simulation of Gas Flows},
  Amsterdam University Press, Amsterdam, Netherlands, 1994.

\bibitem{degond-2010}
P.~Degond, G.~Dimarco, L.~Pareschi, {The moment-guided Monte Carlo method},
  International Journal for Numerical Methods in Fluids 67~(2) (2010) 189--213.

\bibitem{baker-2005B}
L.~L. Baker, N.~G. Hadjiconstantinou, {Variance reduction for Monte Carlo
  solutions of the Boltzmann equation}, Physics of Fluids 17~(5) (2005) 051703.

\bibitem{homolle-2007B}
T.~M. Homolle, N.~G. Hadjiconstantinou, {A low-variance deviational simulation
  Monte Carlo for the Boltzmann equation}, Journal of Computational Physics
  226~(2) (2007) 2341--2358.

\bibitem{Radtke-2011}
G.~Radtke, N.~Hadjiconstantinou, W.~Wagner, {Low-noise Monte Carlo simulation
  of the variable hard sphere gas}, Physics of Fluids 23~(3) (2011) 030606.

\bibitem{pareschi-2000}
L.~Pareschi, G.~Russo, {Asymptotic preserving Monte Carlo methods for the
  Boltzmann equation}, Transport Theory and Statistical Physics 29~(3-5) (2000)
  415--430.

\bibitem{pareschi-2001}
L.~Pareschi, G.~Russo, {Time Relaxed Monte Carlo Methods for the Boltzmann
  Equation}, SIAM Journal on Scientific Computing 23~(4) (2001) 1253--1273.

\bibitem{patronis-2013}
A.~Patronis, D.~A. Lockerby, M.~K. Borg, J.~M. Reese, {Hybrid
  continuum–molecular modelling of multiscale internal gas flows}, Journal of
  Computational Physics 255 (2013) 558--571.

\bibitem{stephani-2013}
K.~Stephani, D.~Goldstein, P.~Varghese, {A non-equilibrium surface reservoir
  approach for hybrid DSMC/Navier–Stokes particle generation}, Journal of
  Computational Physics 232~(1) (2013) 468--481.

\bibitem{xu-2010B}
K.~Xu, J.-C. Huang, {A unified gas-kinetic scheme for continuum and rarefied
  flows}, Journal of Computational Physics 229~(20) (2010) 7747--7764.

\bibitem{huang-2012}
J.-C. Huang, K.~Xu, P.~Yu, {A Unified Gas-Kinetic Scheme for Continuum and
  Rarefied Flows II: Multi-Dimensional Cases}, Communications in Computational
  Physics 12~(3) (2012) 662--690.

\bibitem{Huang-2013}
J.-C. Huang, K.~Xu, P.~Yu, {A Unified Gas-Kinetic Scheme for Continuum and
  Rarefied Flows III: Microflow Simulations}, Communications in Computational
  Physics 14~(5) (2013) 1147–1173.

\bibitem{zhu-2016}
Y.~Zhu, C.~Zhong, K.~Xu, {Implicit unified gas-kinetic scheme for steady state
  solutions in all flow regimes}, Journal of Computational Physics 315 (2016)
  16--38.

\bibitem{SIS}
L.~Wu, J.~Zhang, H.~Liu, Y.-h. Zhang, J.~Reese, {A fast iterative scheme for
  the linearized Boltzmann equation}, Journal of Computational Physics 338
  (2017) 431--451.

\bibitem{CWF}
W.~Su, L.~Zhu, P.~Wang, Y.-h. Zhang, L.~Wu, Can we find steady-state solutions
  to multiscale rarefied gas flows within dozens of iterations?, Journal of
  Computational Physics 407 (2020) 109245.

\bibitem{FCA}
W.~Su, L.~Zhu, L.~Wu, Fast convergence and asymptotic preserving of the general
  synthetic iterative scheme, SIAM Journal on Scientific Computing 42 (2020)
  B1517.

\bibitem{fei-2020}
F.~Fei, J.~Zhang, J.~Li, Z.~Liu, {A unified stochastic particle
  Bhatnagar-Gross-Krook method for multiscale gas flows}, Journal of
  Computational Physics 400 (2020) 108972.

\bibitem{fei-2021}
F.~Fei, P.~Jenny, {A hybrid particle approach based on the unified stochastic
  particle Bhatnagar-Gross-Krook and DSMC methods}, Journal of Computational
  Physics 424 (2021) 109858.

\bibitem{Pfeiffer-2016}
M.~Pfeiffer, H.~Gorji, {Adaptive particle cell algorithm for Fokker–Planck
  based rarefied gas flow simulations}, Computer Physics Communications 213
  (2017) 1--8.

\bibitem{Jenny-2010}
P.~Jenny, M.~Torrilhon, S.~Heinz, A solution algorithm for the fluid dynamic
  equations based on a stochastic model for molecular motion, Journal of
  Computational Physics (2010) 1077--1098.

\bibitem{Gorji-2011}
H.~Gorji, M.~Torrilhon, P.~Jenny, {Fokker–Planck} model for computational
  studies of monatomic rarefied gas flows, Journal of Fluid Mechanics 680
  (2011) 574 -- 601.

\bibitem{Gorji-2014}
H.~Gorji, P.~Jenny, An efficient particle {Fokker-Planck} algorithm for
  rarefied gas flows, Journal of Computational Physics 262 (2014) 325--343.

\bibitem{Bhatnagar-1954}
P.~L. Bhatnagar, E.~P. Gross, M.~Krook, {A Model for Collision Processes in
  Gases. I. Small Amplitude Processes in Charged and Neutral One-Component
  Systems}, Physical Review 94~(3) (1954) 511--525.

\bibitem{Tumuklu-2016}
O.~Tumuklu, Z.~Li, D.~Levin, {Particle Ellipsoidal Statistical
  Bhatnagar–Gross–Krook Approach for Simulation of Hypersonic Shocks}, AIAA
  Journal 54 (2016) 1--16.

\bibitem{Liu-2019}
C.~Liu, Y.~Zhu, K.~Xu, {Unified gas-kinetic wave-particle methods I: Continuum
  and rarefied gas flow}, Journal of Computational Physics 401 (2019) 108977.

\bibitem{Zhu-2019}
Y.~Zhu, C.~Liu, C.~Zhong, K.~Xu, {Unified gas-kinetic wave-particle methods.
  II. Multiscale simulation on unstructured mesh}, Physics of Fluids 31~(6)
  (2019) 067105.

\bibitem{Li-2020}
W.~Li, C.~Liu, Y.~Zhu, J.~Zhang, K.~Xu, {Unified gas-kinetic wave-particle
  methods III: Multiscale photon transport}, Journal of Computational Physics
  408 (2020) 109280.

\bibitem{Radtke-2009}
G.~A. Radtke, N.~G. Hadjiconstantinou, {Variance-reduced particle simulation of
  the Boltzmann transport equation in the relaxation-time approximation},
  Physical Review E 79~(5) (2009) 056711.

\end{thebibliography}

\end{document}